\documentclass[10pt,final,journal]{IEEEtran} 

\usepackage[utf8]{inputenc}
\usepackage{amsmath,color,graphicx,amssymb,mathtools}
\usepackage{ifthen}
\usepackage{textcomp} 
\usepackage{gensymb}
\usepackage{placeins}
\usepackage{float}
\usepackage[bookmarks=false]{hyperref}
\usepackage{tabularx}
\usepackage{epstopdf,multirow}
\usepackage{xcolor}
\usepackage{mathrsfs}
\newcommand{\me}{\mathrm{e}}
\usepackage{tikz,tikz-3dplot}
\usetikzlibrary{decorations.pathreplacing,angles,quotes,arrows.meta}
\usepackage{cite}

\usepackage[T1]{fontenc}

\newcommand{\rhovec}{\boldsymbol{\rho}}
\newcommand{\rhodvec}{\dot{\rhovec}}
\newcommand{\rhoddvec}{\ddot{\rhovec}}
\newcommand{\rhodddvec}{\dddot{\rhovec}}

\begin{document}
\bstctlcite{IEEEexample:BSTcontrol}
\pagenumbering{gobble}

\title{Establishing the Capabilities of the Murchison Widefield Array as a Passive Radar for the Surveillance of Space}
\author{\IEEEauthorblockN{Brendan~Hennessy\IEEEauthorrefmark{1}\IEEEauthorrefmark{2},
Mark~Rutten\IEEEauthorrefmark{3},
Robert~Young\IEEEauthorrefmark{2},
Steven~Tingay\IEEEauthorrefmark{1},
Ashley~Summers\IEEEauthorrefmark{2},
Daniel~Gustainis\IEEEauthorrefmark{2},
Brian~Crosse\IEEEauthorrefmark{1},
Marcin~Sokolowski\IEEEauthorrefmark{1}
}\\
\IEEEauthorblockA{\IEEEauthorrefmark{1}International Centre for Radio Astronomy Research, Curtin University, Bentley, WA 6102, Australia}\\
\IEEEauthorblockA{\IEEEauthorrefmark{2}Defence Science and Technology Group, Edinburgh, SA 5111, Australia.}\\
\IEEEauthorblockA{\IEEEauthorrefmark{3}InTrack Solutions, Adelaide, SA 5000, Australia
\vspace{-1ex}} 
}

\maketitle
\begin{abstract}
This paper describes the use of the Murchison Widefield Array, a low-frequency radio telescope at a radio-quiet Western Australian site, as a radar receiver forming part of a continent-spanning multistatic radar network for the surveillance of space. This paper details the system geometry employed, the orbit-specific radar signal processing, and the orbit determination algorithms necessary to ensure resident space objects are detected, tracked, and propagated. Finally, the paper {includes the results processed after} a short collection campaign utilising several FM radio transmitters across the country, up to a maximum baseline distance of over 2500~km. The results demonstrate the Murchison Widefield Array is able to provide widefield and persistent coverage of objects in low Earth orbit. 
\end{abstract}

\vspace{-2ex}
\section{Introduction}

\IEEEPARstart{R}{adio} telescopes, sensitive instruments designed to detect the faintest signals from the most distant cosmic objects, are proving to be capable receivers for the purpose of radar. Although radio telescope arrays have been used as imaging interferometers, recent interest in phenomena such as pulsars, fast radio bursts, and cosmic rays, is motivating the development of high time-resolution capabilities for interferometric arrays~\cite{tremblay2015high,mcsweeney2020mwa,williamson2021ultra}. These high time-resolution observation modes allow the direct implementation of radar functionality. Thus, researchers are investigating the use of arrays of radio telescopes as radar receivers for the surveillance of space~\cite{7944483,losacco2020initial,dhondea2019mission,9455158}.

This is certainly true for the Murchison Widefield Array (MWA), a low-frequency radio telescope in remote Western Australia, a precursor to the forthcoming Square Kilometre Array (SKA) \cite{2020SPIE11445E..12M}. Even before the MWA commenced operations in 2013, its use as a passive radar for the surveillance of space has been considered and planned~\cite{2006_shi_workshop}. Over the last decade, work has been undertaken to investigate the MWA's ability to detect and track resident space objects (RSOs) in low Earth orbit (LEO) from the reflections of terrestrial transmissions, particularly FM radio~\cite{2013AJ....146..103T,7944483,prabu2020development}. Initial simulations predicted the MWA would be able to detect FM radio reflections from an RSO of a radar cross-section that is as small as 0.5~m$^\text{2}$ at a range of 1000 km~\cite{2013AJ....146..103T}. 

Rather than relying upon pre-existing sources of illumination, many radio telescopes are being utilised as active radar receivers as part of a bistatic system in conjunction with a cooperative radar transmitter~\cite{8741525,dhondea2019mission}. This configuration is beneficial, as the radar transmitter is ideal for the purpose of providing high-powered illumination of the RSOs. However, this is often achieved through the use of narrow-beam transmitter antennas. A direct beam will illuminate only a narrow volume and so the system will only be capable of surveilling a small region, especially when combined with a narrow receiver beam. This is congruous with the operation of the vast majority of space-surveillance radar systems which operate with a single narrow beam, or a fence, configuration~\cite{national2012continuing}.

Passive radar systems are being used for space surveillance purposes with wide-area receiver systems, utilising a (non-cooperative) broad-beam transmitter at a considerable standoff distance.  Such systems include the Low-Frequency Array (LOFAR) radiotelescope, among others~\cite{recent_lofar_paper, sahr1997manastash, 8752314, hennessy2022deployable}. When combined with a receiver with a wide field of regard, a wide area of broadcast transmission ensures a significant volume above the receiver is illuminated. However, this broad coverage also means the relative power directed toward RSOs can be reduced when compared to the cooperative bistatic configuration. Despite this lower directed power, there are benefits to utilising these types of transmitters. A vast coverage volume ensures RSOs may be detected for a significant amount of time, enough to provide accurate orbit estimates from a single pass without any prior knowledge. Additionally, terrestrial broadcasters, such as TV or radio, consist of a distributed network of many transmitters, allowing multiple transmitters to be used as simultaneous sources of radar illumination of the volume above the receiver.

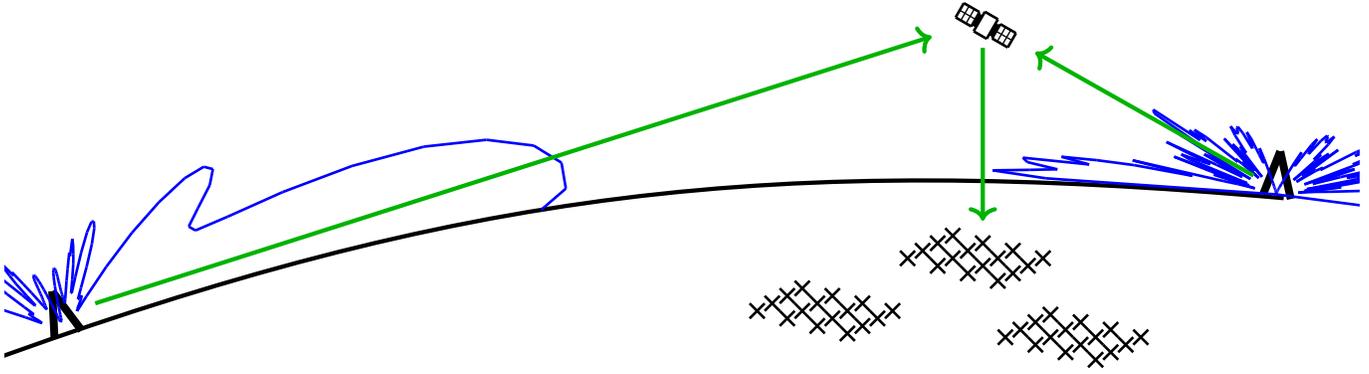
\begin{figure*}[!ht]
\begin{tikzpicture}[dot/.style={draw,fill,circle,inner sep=1pt}]
\clip (-3,-1.5) rectangle + (18,5.5);

    \begin{scope}[xshift=11.2cm, yshift=-1.75cm, rotate=-5]
    \draw[line width=3pt, color=black](2.275, 3.0167) -- (2.45, 3.6);
    \draw[line width=3pt, color=black](2.645, 2.98) -- (2.45, 3.6);
    \end{scope}
    
    \begin{scope}[xshift=-3.48cm, yshift=-4.45cm, rotate=19.485]
    \draw[line width=3pt, color=black](2.275, 3.0167) -- (2.45, 3.6);
    \draw[line width=3pt, color=black](2.645, 2.98) -- (2.45, 3.6);
    \end{scope}

    \draw [black, -, ultra thick] (-3.3,-1.2) to[out=20,in=175] (14,1);

    \begin{scope}[xshift=10.04cm, yshift=3.3cm, rotate=-30, scale=0.1]
    \draw[line width=1pt, color=black](-1, 1.5) -- (1, 1.5); 
    \draw[line width=1pt, color=black](1, 1.5) -- (1, -1.5);
    \draw[line width=1pt, color=black](1, -1.5) -- (-1, -1.5);
    \draw[line width=1pt, color=black](-1, -1.5) -- (-1, 1.5);

    \draw[line width=1pt, color=black](-4, 1.1) -- (-1.6, 1.1); 
    \draw[line width=1pt, color=black](-1.6, 1.1) -- (-1.6, -1.1);
    \draw[line width=1pt, color=black](-1.6, -1.1) -- (-4, -1.1);
    \draw[line width=1pt, color=black](-4, -1.1) -- (-4, 1.1);

    \draw[line width=3pt, color=black](-1.6, 0.4) -- (-1, 0.4);
    \draw[line width=3pt, color=black](-1.6, -0.4) -- (-1, -0.4);

    \draw[line width=3pt, color=black](1.6, 0.4) -- (1, 0.4);
    \draw[line width=3pt, color=black](1.6, -0.4) -- (1, -0.4);

    \draw[line width=1pt, color=black](4, 1.1) -- (1.6, 1.1);
    \draw[line width=1pt, color=black](1.6, 1.1) -- (1.6, -1.1);
    \draw[line width=1pt, color=black](1.6, -1.1) -- (4, -1.1);
    \draw[line width=1pt, color=black](4, -1.1) -- (4, 1.1);

    \draw[line width=0.5pt, color=black](1.6, 0) -- (4, 0);
    \draw[line width=0.5pt, color=black](2.4, -1.1) -- (2.4, 1.1);
    \draw[line width=0.5pt, color=black](3.2, -1.1) -- (3.2, 1.1);

    \draw[line width=0.5pt, color=black](-1.6, 0) -- (-4, 0);
    \draw[line width=0.5pt, color=black](-2.4, -1.1) -- (-2.4, 1.1);
    \draw[line width=0.5pt, color=black](-3.2, -1.1) -- (-3.2, 1.1);

    \end{scope}

    \begin{scope}[xshift=7cm, yshift=-0.5cm]
        \begin{scope}[xshift=0, yshift=0]
            \begin{scope}[xshift=0, yshift=0]
            \draw[line width=1pt, color=black](-.1, -.1) -- (.1, .1);
            \draw[line width=1pt, color=black](-.1, .1) -- (.1, -.1);
            \end{scope}
            \begin{scope}[xshift=0.2cm, yshift=0.1cm]
            \draw[line width=1pt, color=black](-.1, -.1) -- (.1, .1);
            \draw[line width=1pt, color=black](-.1, .1) -- (.1, -.1);
            \end{scope}
            \begin{scope}[xshift=0.4cm, yshift=0.2cm]
            \draw[line width=1pt, color=black](-.1, -.1) -- (.1, .1);
            \draw[line width=1pt, color=black](-.1, .1) -- (.1, -.1);
            \end{scope}
            \begin{scope}[xshift=0.6cm, yshift=0.3cm]
            \draw[line width=1pt, color=black](-.1, -.1) -- (.1, .1);
            \draw[line width=1pt, color=black](-.1, .1) -- (.1, -.1);
            \end{scope}
        \end{scope}
        \begin{scope}[xshift=0.4cm, yshift=-0.1cm]
            \begin{scope}[xshift=0, yshift=0]
            \draw[line width=1pt, color=black](-.1, -.1) -- (.1, .1);
            \draw[line width=1pt, color=black](-.1, .1) -- (.1, -.1);
            \end{scope}
            \begin{scope}[xshift=0.2cm, yshift=0.1cm]
            \draw[line width=1pt, color=black](-.1, -.1) -- (.1, .1);
            \draw[line width=1pt, color=black](-.1, .1) -- (.1, -.1);
            \end{scope}
            \begin{scope}[xshift=0.4cm, yshift=0.2cm]
            \draw[line width=1pt, color=black](-.1, -.1) -- (.1, .1);
            \draw[line width=1pt, color=black](-.1, .1) -- (.1, -.1);
            \end{scope}
            \begin{scope}[xshift=0.6cm, yshift=0.3cm]
            \draw[line width=1pt, color=black](-.1, -.1) -- (.1, .1);
            \draw[line width=1pt, color=black](-.1, .1) -- (.1, -.1);
            \end{scope}
        \end{scope}
        \begin{scope}[xshift=0.8cm, yshift=-0.2cm]
            \begin{scope}[xshift=0, yshift=0]
            \draw[line width=1pt, color=black](-.1, -.1) -- (.1, .1);
            \draw[line width=1pt, color=black](-.1, .1) -- (.1, -.1);
            \end{scope}
            \begin{scope}[xshift=0.2cm, yshift=0.1cm]
            \draw[line width=1pt, color=black](-.1, -.1) -- (.1, .1);
            \draw[line width=1pt, color=black](-.1, .1) -- (.1, -.1);
            \end{scope}
            \begin{scope}[xshift=0.4cm, yshift=0.2cm]
            \draw[line width=1pt, color=black](-.1, -.1) -- (.1, .1);
            \draw[line width=1pt, color=black](-.1, .1) -- (.1, -.1);
            \end{scope}
            \begin{scope}[xshift=0.6cm, yshift=0.3cm]
            \draw[line width=1pt, color=black](-.1, -.1) -- (.1, .1);
            \draw[line width=1pt, color=black](-.1, .1) -- (.1, -.1);
            \end{scope}
        \end{scope}
        \begin{scope}[xshift=1.2cm, yshift=-0.3cm]
            \begin{scope}[xshift=0, yshift=0]
            \draw[line width=1pt, color=black](-.1, -.1) -- (.1, .1);
            \draw[line width=1pt, color=black](-.1, .1) -- (.1, -.1);
            \end{scope}
            \begin{scope}[xshift=0.2cm, yshift=0.1cm]
            \draw[line width=1pt, color=black](-.1, -.1) -- (.1, .1);
            \draw[line width=1pt, color=black](-.1, .1) -- (.1, -.1);
            \end{scope}
            \begin{scope}[xshift=0.4cm, yshift=0.2cm]
            \draw[line width=1pt, color=black](-.1, -.1) -- (.1, .1);
            \draw[line width=1pt, color=black](-.1, .1) -- (.1, -.1);
            \end{scope}
            \begin{scope}[xshift=0.6cm, yshift=0.3cm]
            \draw[line width=1pt, color=black](-.1, -.1) -- (.1, .1);
            \draw[line width=1pt, color=black](-.1, .1) -- (.1, -.1);
            \end{scope}
        \end{scope}        
    \end{scope}

    \begin{scope}[xshift=9cm, yshift=0.2cm]
        \begin{scope}[xshift=0, yshift=0]
            \begin{scope}[xshift=0, yshift=0]
            \draw[line width=1pt, color=black](-.1, -.1) -- (.1, .1);
            \draw[line width=1pt, color=black](-.1, .1) -- (.1, -.1);
            \end{scope}
            \begin{scope}[xshift=0.2cm, yshift=0.1cm]
            \draw[line width=1pt, color=black](-.1, -.1) -- (.1, .1);
            \draw[line width=1pt, color=black](-.1, .1) -- (.1, -.1);
            \end{scope}
            \begin{scope}[xshift=0.4cm, yshift=0.2cm]
            \draw[line width=1pt, color=black](-.1, -.1) -- (.1, .1);
            \draw[line width=1pt, color=black](-.1, .1) -- (.1, -.1);
            \end{scope}
            \begin{scope}[xshift=0.6cm, yshift=0.3cm]
            \draw[line width=1pt, color=black](-.1, -.1) -- (.1, .1);
            \draw[line width=1pt, color=black](-.1, .1) -- (.1, -.1);
            \end{scope}
        \end{scope}
        \begin{scope}[xshift=0.4cm, yshift=-0.1cm]
            \begin{scope}[xshift=0, yshift=0]
            \draw[line width=1pt, color=black](-.1, -.1) -- (.1, .1);
            \draw[line width=1pt, color=black](-.1, .1) -- (.1, -.1);
            \end{scope}
            \begin{scope}[xshift=0.2cm, yshift=0.1cm]
            \draw[line width=1pt, color=black](-.1, -.1) -- (.1, .1);
            \draw[line width=1pt, color=black](-.1, .1) -- (.1, -.1);
            \end{scope}
            \begin{scope}[xshift=0.4cm, yshift=0.2cm]
            \draw[line width=1pt, color=black](-.1, -.1) -- (.1, .1);
            \draw[line width=1pt, color=black](-.1, .1) -- (.1, -.1);
            \end{scope}
            \begin{scope}[xshift=0.6cm, yshift=0.3cm]
            \draw[line width=1pt, color=black](-.1, -.1) -- (.1, .1);
            \draw[line width=1pt, color=black](-.1, .1) -- (.1, -.1);
            \end{scope}
        \end{scope}
        \begin{scope}[xshift=0.8cm, yshift=-0.2cm]
            \begin{scope}[xshift=0, yshift=0]
            \draw[line width=1pt, color=black](-.1, -.1) -- (.1, .1);
            \draw[line width=1pt, color=black](-.1, .1) -- (.1, -.1);
            \end{scope}
            \begin{scope}[xshift=0.2cm, yshift=0.1cm]
            \draw[line width=1pt, color=black](-.1, -.1) -- (.1, .1);
            \draw[line width=1pt, color=black](-.1, .1) -- (.1, -.1);
            \end{scope}
            \begin{scope}[xshift=0.4cm, yshift=0.2cm]
            \draw[line width=1pt, color=black](-.1, -.1) -- (.1, .1);
            \draw[line width=1pt, color=black](-.1, .1) -- (.1, -.1);
            \end{scope}
            \begin{scope}[xshift=0.6cm, yshift=0.3cm]
            \draw[line width=1pt, color=black](-.1, -.1) -- (.1, .1);
            \draw[line width=1pt, color=black](-.1, .1) -- (.1, -.1);
            \end{scope}
        \end{scope}
        \begin{scope}[xshift=1.2cm, yshift=-0.3cm]
            \begin{scope}[xshift=0, yshift=0]
            \draw[line width=1pt, color=black](-.1, -.1) -- (.1, .1);
            \draw[line width=1pt, color=black](-.1, .1) -- (.1, -.1);
            \end{scope}
            \begin{scope}[xshift=0.2cm, yshift=0.1cm]
            \draw[line width=1pt, color=black](-.1, -.1) -- (.1, .1);
            \draw[line width=1pt, color=black](-.1, .1) -- (.1, -.1);
            \end{scope}
            \begin{scope}[xshift=0.4cm, yshift=0.2cm]
            \draw[line width=1pt, color=black](-.1, -.1) -- (.1, .1);
            \draw[line width=1pt, color=black](-.1, .1) -- (.1, -.1);
            \end{scope}
            \begin{scope}[xshift=0.6cm, yshift=0.3cm]
            \draw[line width=1pt, color=black](-.1, -.1) -- (.1, .1);
            \draw[line width=1pt, color=black](-.1, .1) -- (.1, -.1);
            \end{scope}
        \end{scope}        
    \end{scope}

    \begin{scope}[xshift=10.3cm, yshift=-0.85cm]
        \begin{scope}[xshift=0, yshift=0]
            \begin{scope}[xshift=0, yshift=0]
            \draw[line width=1pt, color=black](-.1, -.1) -- (.1, .1);
            \draw[line width=1pt, color=black](-.1, .1) -- (.1, -.1);
            \end{scope}
            \begin{scope}[xshift=0.2cm, yshift=0.1cm]
            \draw[line width=1pt, color=black](-.1, -.1) -- (.1, .1);
            \draw[line width=1pt, color=black](-.1, .1) -- (.1, -.1);
            \end{scope}
            \begin{scope}[xshift=0.4cm, yshift=0.2cm]
            \draw[line width=1pt, color=black](-.1, -.1) -- (.1, .1);
            \draw[line width=1pt, color=black](-.1, .1) -- (.1, -.1);
            \end{scope}
            \begin{scope}[xshift=0.6cm, yshift=0.3cm]
            \draw[line width=1pt, color=black](-.1, -.1) -- (.1, .1);
            \draw[line width=1pt, color=black](-.1, .1) -- (.1, -.1);
            \end{scope}
        \end{scope}
        \begin{scope}[xshift=0.4cm, yshift=-0.1cm]
            \begin{scope}[xshift=0, yshift=0]
            \draw[line width=1pt, color=black](-.1, -.1) -- (.1, .1);
            \draw[line width=1pt, color=black](-.1, .1) -- (.1, -.1);
            \end{scope}
            \begin{scope}[xshift=0.2cm, yshift=0.1cm]
            \draw[line width=1pt, color=black](-.1, -.1) -- (.1, .1);
            \draw[line width=1pt, color=black](-.1, .1) -- (.1, -.1);
            \end{scope}
            \begin{scope}[xshift=0.4cm, yshift=0.2cm]
            \draw[line width=1pt, color=black](-.1, -.1) -- (.1, .1);
            \draw[line width=1pt, color=black](-.1, .1) -- (.1, -.1);
            \end{scope}
            \begin{scope}[xshift=0.6cm, yshift=0.3cm]
            \draw[line width=1pt, color=black](-.1, -.1) -- (.1, .1);
            \draw[line width=1pt, color=black](-.1, .1) -- (.1, -.1);
            \end{scope}
        \end{scope}
        \begin{scope}[xshift=0.8cm, yshift=-0.2cm]
            \begin{scope}[xshift=0, yshift=0]
            \draw[line width=1pt, color=black](-.1, -.1) -- (.1, .1);
            \draw[line width=1pt, color=black](-.1, .1) -- (.1, -.1);
            \end{scope}
            \begin{scope}[xshift=0.2cm, yshift=0.1cm]
            \draw[line width=1pt, color=black](-.1, -.1) -- (.1, .1);
            \draw[line width=1pt, color=black](-.1, .1) -- (.1, -.1);
            \end{scope}
            \begin{scope}[xshift=0.4cm, yshift=0.2cm]
            \draw[line width=1pt, color=black](-.1, -.1) -- (.1, .1);
            \draw[line width=1pt, color=black](-.1, .1) -- (.1, -.1);
            \end{scope}
            \begin{scope}[xshift=0.6cm, yshift=0.3cm]
            \draw[line width=1pt, color=black](-.1, -.1) -- (.1, .1);
            \draw[line width=1pt, color=black](-.1, .1) -- (.1, -.1);
            \end{scope}
        \end{scope}
        \begin{scope}[xshift=1.2cm, yshift=-0.3cm]
            \begin{scope}[xshift=0, yshift=0]
            \draw[line width=1pt, color=black](-.1, -.1) -- (.1, .1);
            \draw[line width=1pt, color=black](-.1, .1) -- (.1, -.1);
            \end{scope}
            \begin{scope}[xshift=0.2cm, yshift=0.1cm]
            \draw[line width=1pt, color=black](-.1, -.1) -- (.1, .1);
            \draw[line width=1pt, color=black](-.1, .1) -- (.1, -.1);
            \end{scope}
            \begin{scope}[xshift=0.4cm, yshift=0.2cm]
            \draw[line width=1pt, color=black](-.1, -.1) -- (.1, .1);
            \draw[line width=1pt, color=black](-.1, .1) -- (.1, -.1);
            \end{scope}
            \begin{scope}[xshift=0.6cm, yshift=0.3cm]
            \draw[line width=1pt, color=black](-.1, -.1) -- (.1, .1);
            \draw[line width=1pt, color=black](-.1, .1) -- (.1, -.1);
            \end{scope}
        \end{scope}        
    \end{scope}

    \begin{scope}[xshift=-2.2cm,yshift=-0.8cm,rotate=19,xscale=3.5,yscale=10]
        \draw[line width=1pt, color=blue](1.862865, -0.05053) -- (1.977699, -0.034521);
        \draw[line width=1pt, color=blue](1.977699, -0.034521) -- (1.994000, 0.000000);
        \draw[line width=1pt, color=blue](1.994000, 0.000000) -- (1.915708, 0.033439);
        \draw[line width=1pt, color=blue](1.915708, 0.033439) -- (1.752932, 0.061214);
        \draw[line width=1pt, color=blue](1.752932, 0.061214) -- (1.519914, 0.079655);
        \draw[line width=1pt, color=blue](1.519914, 0.079655) -- (1.236979, 0.086498);
        \draw[line width=1pt, color=blue](1.236979, 0.086498) -- (0.956347, 0.083670);
        \draw[line width=1pt, color=blue](0.956347, 0.083670) -- (0.727990, 0.076515);
        \draw[line width=1pt, color=blue](0.727990, 0.076515) -- (0.595528, 0.073122);
        \draw[line width=1pt, color=blue](0.595528, 0.073122) -- (0.576336, 0.080999);
        \draw[line width=1pt, color=blue](0.576336, 0.080999) -- (0.632121, 0.100118);
        \draw[line width=1pt, color=blue](0.632121, 0.100118) -- (0.703153, 0.123985);
        \draw[line width=1pt, color=blue](0.703153, 0.123985) -- (0.734257, 0.142725);
        \draw[line width=1pt, color=blue](0.734257, 0.142725) -- (0.704266, 0.149696);
        \draw[line width=1pt, color=blue](0.704266, 0.149696) -- (0.621648, 0.143519);
        \draw[line width=1pt, color=blue](0.621648, 0.143519) -- (0.498732, 0.124348);
        \draw[line width=1pt, color=blue](0.498732, 0.124348) -- (0.365120, 0.097834);
        \draw[line width=1pt, color=blue](0.365120, 0.097834) -- (0.230703, 0.066153);
        \draw[line width=1pt, color=blue](0.230703, 0.066153) -- (0.122407, 0.037424);
        \draw[line width=1pt, color=blue](0.122407, 0.037424) -- (0.070378, 0.022867);
        \draw[line width=1pt, color=blue](0.070378, 0.022867) -- (0.073750, 0.025394);
        \draw[line width=1pt, color=blue](0.073750, 0.025394) -- (0.099607, 0.036254);
        \draw[line width=1pt, color=blue](0.099607, 0.036254) -- (0.106428, 0.040854);
        \draw[line width=1pt, color=blue](0.106428, 0.040854) -- (0.092718, 0.037461);
        \draw[line width=1pt, color=blue](0.092718, 0.037461) -- (0.088368, 0.037510);
        \draw[line width=1pt, color=blue](0.088368, 0.037510) -- (0.113280, 0.050435);
        \draw[line width=1pt, color=blue](0.113280, 0.050435) -- (0.159510, 0.074381);
        \draw[line width=1pt, color=blue](0.159510, 0.074381) -- (0.206723, 0.100825);
        \draw[line width=1pt, color=blue](0.206723, 0.100825) -- (0.235226, 0.119853);
        \draw[line width=1pt, color=blue](0.235226, 0.119853) -- (0.241928, 0.128635);
        \draw[line width=1pt, color=blue](0.241928, 0.128635) -- (0.232649, 0.128959);
        \draw[line width=1pt, color=blue](0.232649, 0.128959) -- (0.204382, 0.118000);
        \draw[line width=1pt, color=blue](0.204382, 0.118000) -- (0.157719, 0.094767);
        \draw[line width=1pt, color=blue](0.157719, 0.094767) -- (0.106854, 0.066770);
        \draw[line width=1pt, color=blue](0.106854, 0.066770) -- (0.073803, 0.047928);
        \draw[line width=1pt, color=blue](0.073803, 0.047928) -- (0.071297, 0.048091);
        \draw[line width=1pt, color=blue](0.071297, 0.048091) -- (0.091745, 0.064241);
        \draw[line width=1pt, color=blue](0.091745, 0.064241) -- (0.114880, 0.083466);
        \draw[line width=1pt, color=blue](0.114880, 0.083466) -- (0.127782, 0.096290);
        \draw[line width=1pt, color=blue](0.127782, 0.096290) -- (0.138690, 0.108356);
        \draw[line width=1pt, color=blue](0.138690, 0.108356) -- (0.142995, 0.115795);
        \draw[line width=1pt, color=blue](0.142995, 0.115795) -- (0.113375, 0.095133);
        \draw[line width=1pt, color=blue](0.113375, 0.095133) -- (0.114716, 0.099721);
        \draw[line width=1pt, color=blue](0.114716, 0.099721) -- (0.089177, 0.080296);
        \draw[line width=1pt, color=blue](0.089177, 0.080296) -- (0.070210, 0.065472);
        \draw[line width=1pt, color=blue](0.070210, 0.065472) -- (0.054670, 0.052794);
        \draw[line width=1pt, color=blue](0.054670, 0.052794) -- (0.043841, 0.043841);
        \draw[line width=1pt, color=blue](0.043841, 0.043841) -- (0.037512, 0.038844);
        \draw[line width=1pt, color=blue](0.037512, 0.038844) -- (0.035464, 0.038030);
        \draw[line width=1pt, color=blue](0.035464, 0.038030) -- (0.036133, 0.040130);
        \draw[line width=1pt, color=blue](0.036133, 0.040130) -- (0.036739, 0.042264);
        \draw[line width=1pt, color=blue](0.036739, 0.042264) -- (0.035996, 0.042898);
        \draw[line width=1pt, color=blue](0.035996, 0.042898) -- (0.036501, 0.045074);
        \draw[line width=1pt, color=blue](0.036501, 0.045074) -- (0.036940, 0.047281);
        \draw[line width=1pt, color=blue](0.036940, 0.047281) -- (0.037313, 0.049515);
        \draw[line width=1pt, color=blue](0.037313, 0.049515) -- (0.037618, 0.051777);
        \draw[line width=1pt, color=blue](0.037618, 0.051777) -- (0.037856, 0.054064);
        \draw[line width=1pt, color=blue](0.037856, 0.054064) -- (0.038025, 0.056375);
        \draw[line width=1pt, color=blue](0.038025, 0.056375) -- (0.039214, 0.060384);
        \draw[line width=1pt, color=blue](0.039214, 0.060384) -- (0.040274, 0.064452);
        \draw[line width=1pt, color=blue](0.040274, 0.064452) -- (0.041203, 0.068573);
        \draw[line width=1pt, color=blue](0.041203, 0.068573) -- (0.043000, 0.074478);
        \draw[line width=1pt, color=blue](0.043000, 0.074478) -- (0.043633, 0.078716);
        \draw[line width=1pt, color=blue](0.043633, 0.078716) -- (0.044130, 0.082997);
        \draw[line width=1pt, color=blue](0.044130, 0.082997) -- (0.043583, 0.085537);
        \draw[line width=1pt, color=blue](0.043583, 0.085537) -- (0.042084, 0.086284);
        \draw[line width=1pt, color=blue](0.042084, 0.086284) -- (0.040571, 0.087006);
        \draw[line width=1pt, color=blue](0.040571, 0.087006) -- (0.038233, 0.085873);
        \draw[line width=1pt, color=blue](0.038233, 0.085873) -- (0.035947, 0.084686);
        \draw[line width=1pt, color=blue](0.035947, 0.084686) -- (0.032965, 0.081592);
        \draw[line width=1pt, color=blue](0.032965, 0.081592) -- (0.029386, 0.076554);
        \draw[line width=1pt, color=blue](0.029386, 0.076554) -- (0.025994, 0.071417);
        \draw[line width=1pt, color=blue](0.025994, 0.071417) -- (0.022139, 0.064295);
        \draw[line width=1pt, color=blue](0.022139, 0.064295) -- (0.018541, 0.057063);
        \draw[line width=1pt, color=blue](0.018541, 0.057063) -- (0.015203, 0.049728);
        \draw[line width=1pt, color=blue](0.015203, 0.049728) -- (0.012128, 0.042296);
        \draw[line width=1pt, color=blue](0.012128, 0.042296) -- (0.009317, 0.034773);
        \draw[line width=1pt, color=blue](0.009317, 0.034773) -- (0.007258, 0.029109);
        \draw[line width=1pt, color=blue](0.007258, 0.029109) -- (0.005849, 0.025334);
        \draw[line width=1pt, color=blue](0.005849, 0.025334) -- (0.004574, 0.021519);
        \draw[line width=1pt, color=blue](0.004574, 0.021519) -- (0.003816, 0.019633);
        \draw[line width=1pt, color=blue](0.003816, 0.019633) -- (0.003126, 0.017727);
        \draw[line width=1pt, color=blue](0.003126, 0.017727) -- (0.002816, 0.017778);
        \draw[line width=1pt, color=blue](0.002816, 0.017778) -- (0.002505, 0.017825);
        \draw[line width=1pt, color=blue](0.002505, 0.017825) -- (0.002194, 0.017866);
        \draw[line width=1pt, color=blue](0.002194, 0.017866) -- (0.001882, 0.017901);
        \draw[line width=1pt, color=blue](0.001882, 0.017901) -- (0.001569, 0.017932);
        \draw[line width=1pt, color=blue](0.001569, 0.017932) -- (0.001256, 0.017956);
        \draw[line width=1pt, color=blue](0.001256, 0.017956) -- (0.000942, 0.017975);
        \draw[line width=1pt, color=blue](0.000942, 0.017975) -- (0.000558, 0.015990);
        \draw[line width=1pt, color=blue](0.000558, 0.015990) -- (0.000279, 0.015998);
        \draw[line width=1pt, color=blue](0.000279, 0.015998) -- (0.000000, 0.016000);
        \draw[line width=1pt, color=blue](0.000000, 0.016000) -- (-0.000279, 0.015998);
        \draw[line width=1pt, color=blue](-0.000279, 0.015998) -- (-0.000558, 0.015990);
        \draw[line width=1pt, color=blue](-0.000558, 0.015990) -- (-0.000942, 0.017975);
        \draw[line width=1pt, color=blue](-0.000942, 0.017975) -- (-0.001256, 0.017956);
        \draw[line width=1pt, color=blue](-0.001256, 0.017956) -- (-0.001569, 0.017932);
        \draw[line width=1pt, color=blue](-0.001569, 0.017932) -- (-0.001882, 0.017901);
        \draw[line width=1pt, color=blue](-0.001882, 0.017901) -- (-0.002194, 0.017866);
        \draw[line width=1pt, color=blue](-0.002194, 0.017866) -- (-0.002505, 0.017825);
        \draw[line width=1pt, color=blue](-0.002505, 0.017825) -- (-0.002816, 0.017778);
        \draw[line width=1pt, color=blue](-0.002816, 0.017778) -- (-0.003126, 0.017727);
        \draw[line width=1pt, color=blue](-0.003126, 0.017727) -- (-0.003816, 0.019633);
        \draw[line width=1pt, color=blue](-0.003816, 0.019633) -- (-0.004574, 0.021519);
        \draw[line width=1pt, color=blue](-0.004574, 0.021519) -- (-0.005849, 0.025334);
        \draw[line width=1pt, color=blue](-0.005849, 0.025334) -- (-0.007258, 0.029109);
        \draw[line width=1pt, color=blue](-0.007258, 0.029109) -- (-0.009317, 0.034773);
        \draw[line width=1pt, color=blue](-0.009317, 0.034773) -- (-0.012128, 0.042296);
        \draw[line width=1pt, color=blue](-0.012128, 0.042296) -- (-0.015203, 0.049728);
        \draw[line width=1pt, color=blue](-0.015203, 0.049728) -- (-0.018541, 0.057063);
        \draw[line width=1pt, color=blue](-0.018541, 0.057063) -- (-0.022139, 0.064295);
        \draw[line width=1pt, color=blue](-0.022139, 0.064295) -- (-0.025994, 0.071417);
        \draw[line width=1pt, color=blue](-0.025994, 0.071417) -- (-0.029386, 0.076554);
        \draw[line width=1pt, color=blue](-0.029386, 0.076554) -- (-0.032965, 0.081592);
        \draw[line width=1pt, color=blue](-0.032965, 0.081592) -- (-0.035947, 0.084686);
        \draw[line width=1pt, color=blue](-0.035947, 0.084686) -- (-0.038233, 0.085873);
        \draw[line width=1pt, color=blue](-0.038233, 0.085873) -- (-0.040571, 0.087006);
        \draw[line width=1pt, color=blue](-0.040571, 0.087006) -- (-0.042084, 0.086284);
        \draw[line width=1pt, color=blue](-0.042084, 0.086284) -- (-0.043583, 0.085537);
        \draw[line width=1pt, color=blue](-0.043583, 0.085537) -- (-0.044130, 0.082997);
        \draw[line width=1pt, color=blue](-0.044130, 0.082997) -- (-0.043633, 0.078716);
        \draw[line width=1pt, color=blue](-0.043633, 0.078716) -- (-0.043000, 0.074478);
        \draw[line width=1pt, color=blue](-0.043000, 0.074478) -- (-0.041203, 0.068573);
        \draw[line width=1pt, color=blue](-0.041203, 0.068573) -- (-0.040274, 0.064452);
        \draw[line width=1pt, color=blue](-0.040274, 0.064452) -- (-0.039214, 0.060384);
        \draw[line width=1pt, color=blue](-0.039214, 0.060384) -- (-0.038025, 0.056375);
        \draw[line width=1pt, color=blue](-0.038025, 0.056375) -- (-0.037856, 0.054064);
        \draw[line width=1pt, color=blue](-0.037856, 0.054064) -- (-0.037618, 0.051777);
        \draw[line width=1pt, color=blue](-0.037618, 0.051777) -- (-0.037313, 0.049515);
        \draw[line width=1pt, color=blue](-0.037313, 0.049515) -- (-0.036940, 0.047281);
        \draw[line width=1pt, color=blue](-0.036940, 0.047281) -- (-0.036501, 0.045074);
        \draw[line width=1pt, color=blue](-0.036501, 0.045074) -- (-0.035996, 0.042898);
        \draw[line width=1pt, color=blue](-0.035996, 0.042898) -- (-0.036739, 0.042264);
        \draw[line width=1pt, color=blue](-0.036739, 0.042264) -- (-0.036133, 0.040130);
        \draw[line width=1pt, color=blue](-0.036133, 0.040130) -- (-0.035464, 0.038030);
        \draw[line width=1pt, color=blue](-0.035464, 0.038030) -- (-0.037512, 0.038844);
        \draw[line width=1pt, color=blue](-0.037512, 0.038844) -- (-0.043841, 0.043841);
        \draw[line width=1pt, color=blue](-0.043841, 0.043841) -- (-0.054670, 0.052794);
        \draw[line width=1pt, color=blue](-0.054670, 0.052794) -- (-0.070210, 0.065472);
        \draw[line width=1pt, color=blue](-0.070210, 0.065472) -- (-0.089177, 0.080296);
        \draw[line width=1pt, color=blue](-0.089177, 0.080296) -- (-0.114716, 0.099721);
        \draw[line width=1pt, color=blue](-0.114716, 0.099721) -- (-0.113375, 0.095133);
        \draw[line width=1pt, color=blue](-0.113375, 0.095133) -- (-0.142995, 0.115795);
        \draw[line width=1pt, color=blue](-0.142995, 0.115795) -- (-0.138690, 0.108356);
        \draw[line width=1pt, color=blue](-0.138690, 0.108356) -- (-0.127782, 0.096290);
        \draw[line width=1pt, color=blue](-0.127782, 0.096290) -- (-0.114880, 0.083466);
        \draw[line width=1pt, color=blue](-0.114880, 0.083466) -- (-0.091745, 0.064241);
        \draw[line width=1pt, color=blue](-0.091745, 0.064241) -- (-0.071297, 0.048091);
        \draw[line width=1pt, color=blue](-0.071297, 0.048091) -- (-0.073803, 0.047928);
        \draw[line width=1pt, color=blue](-0.073803, 0.047928) -- (-0.106854, 0.066770);
        \draw[line width=1pt, color=blue](-0.106854, 0.066770) -- (-0.157719, 0.094767);
        \draw[line width=1pt, color=blue](-0.157719, 0.094767) -- (-0.204382, 0.118000);
        \draw[line width=1pt, color=blue](-0.204382, 0.118000) -- (-0.232649, 0.128959);
        \draw[line width=1pt, color=blue](-0.232649, 0.128959) -- (-0.241928, 0.128635);
        \draw[line width=1pt, color=blue](-0.241928, 0.128635) -- (-0.235226, 0.119853);
        \draw[line width=1pt, color=blue](-0.235226, 0.119853) -- (-0.206723, 0.100825);
        \draw[line width=1pt, color=blue](-0.206723, 0.100825) -- (-0.159510, 0.074381);
        \draw[line width=1pt, color=blue](-0.159510, 0.074381) -- (-0.113280, 0.050435);
        \draw[line width=1pt, color=blue](-0.113280, 0.050435) -- (-0.088368, 0.037510);
        \draw[line width=1pt, color=blue](-0.088368, 0.037510) -- (-0.092718, 0.037461);
        \draw[line width=1pt, color=blue](-0.092718, 0.037461) -- (-0.106428, 0.040854);
        \draw[line width=1pt, color=blue](-0.106428, 0.040854) -- (-0.099607, 0.036254);
        \draw[line width=1pt, color=blue](-0.099607, 0.036254) -- (-0.073750, 0.025394);
        \draw[line width=1pt, color=blue](-0.073750, 0.025394) -- (-0.070378, 0.022867);
        \draw[line width=1pt, color=blue](-0.070378, 0.022867) -- (-0.122407, 0.037424);
        \draw[line width=1pt, color=blue](-0.122407, 0.037424) -- (-0.230703, 0.066153);
        \draw[line width=1pt, color=blue](-0.230703, 0.066153) -- (-0.365120, 0.097834);
        \draw[line width=1pt, color=blue](-0.365120, 0.097834) -- (-0.498732, 0.124348);
        \draw[line width=1pt, color=blue](-0.498732, 0.124348) -- (-0.621648, 0.143519);
        \draw[line width=1pt, color=blue](-0.621648, 0.143519) -- (-0.704266, 0.149696);
        \draw[line width=1pt, color=blue](-0.704266, 0.149696) -- (-0.734257, 0.142725);
        \draw[line width=1pt, color=blue](-0.734257, 0.142725) -- (-0.703153, 0.123985);
        \draw[line width=1pt, color=blue](-0.703153, 0.123985) -- (-0.632121, 0.100118);
        \draw[line width=1pt, color=blue](-0.632121, 0.100118) -- (-0.576336, 0.080999);
        \draw[line width=1pt, color=blue](-0.576336, 0.080999) -- (-0.595528, 0.073122);
        \draw[line width=1pt, color=blue](-0.595528, 0.073122) -- (-0.727990, 0.076515);
        \draw[line width=1pt, color=blue](-0.727990, 0.076515) -- (-0.956347, 0.083670);
        \draw[line width=1pt, color=blue](-0.956347, 0.083670) -- (-1.236979, 0.086498);
        \draw[line width=1pt, color=blue](-1.236979, 0.086498) -- (-1.519914, 0.079655);
        \draw[line width=1pt, color=blue](-1.519914, 0.079655) -- (-1.752932, 0.061214);
        \draw[line width=1pt, color=blue](-1.752932, 0.061214) -- (-1.915708, 0.033439);
        \draw[line width=1pt, color=blue](-1.915708, 0.033439) -- (-1.994000, 0.000000);
        \draw[line width=1pt, color=blue](-1.994000, 0.000000) -- (-1.977699, -0.034521);
        \draw[line width=1pt, color=blue](-1.977699, -0.034521) -- (-1.862865, -0.065053);
    \end{scope}
    
    \begin{scope}[xshift=13.9cm,yshift=1.05cm,rotate=-5.7,xscale=2,yscale=1.7]
    
\draw[line width=1pt, color=blue](0.000000, -0.000000) -- (1.533066, -0.053536);
\draw[line width=1pt, color=blue](1.533066, -0.053536) -- (1.891712, -0.033020);
\draw[line width=1pt, color=blue](1.891712, -0.033020) -- (1.644000, 0.000000);
\draw[line width=1pt, color=blue](1.644000, 0.000000) -- (1.463777, 0.025550);
\draw[line width=1pt, color=blue](1.463777, 0.025550) -- (1.666984, 0.058212);
\draw[line width=1pt, color=blue](1.666984, 0.058212) -- (1.699667, 0.089076);
\draw[line width=1pt, color=blue](1.699667, 0.089076) -- (1.256931, 0.087893);
\draw[line width=1pt, color=blue](1.256931, 0.087893) -- (1.444482, 0.126376);
\draw[line width=1pt, color=blue](1.444482, 0.126376) -- (1.209339, 0.127107);
\draw[line width=1pt, color=blue](1.209339, 0.127107) -- (0.958800, 0.117726);
\draw[line width=1pt, color=blue](0.958800, 0.117726) -- (0.566433, 0.079607);
\draw[line width=1pt, color=blue](0.566433, 0.079607) -- (0.971885, 0.153932);
\draw[line width=1pt, color=blue](0.971885, 0.153932) -- (0.352561, 0.062166);
\draw[line width=1pt, color=blue](0.352561, 0.062166) -- (0.261113, 0.050755);
\draw[line width=1pt, color=blue](0.261113, 0.050755) -- (0.144766, 0.030771);
\draw[line width=1pt, color=blue](0.144766, 0.030771) -- (0.233849, 0.053988);
\draw[line width=1pt, color=blue](0.233849, 0.053988) -- (0.424990, 0.105962);
\draw[line width=1pt, color=blue](0.424990, 0.105962) -- (0.185458, 0.049693);
\draw[line width=1pt, color=blue](0.185458, 0.049693) -- (0.446025, 0.127896);
\draw[line width=1pt, color=blue](0.446025, 0.127896) -- (0.700015, 0.214016);
\draw[line width=1pt, color=blue](0.700015, 0.214016) -- (0.199722, 0.064894);
\draw[line width=1pt, color=blue](0.199722, 0.064894) -- (0.593786, 0.204457);
\draw[line width=1pt, color=blue](0.593786, 0.204457) -- (0.654026, 0.238046);
\draw[line width=1pt, color=blue](0.654026, 0.238046) -- (0.517204, 0.198536);
\draw[line width=1pt, color=blue](0.517204, 0.198536) -- (0.456174, 0.184306);
\draw[line width=1pt, color=blue](0.456174, 0.184306) -- (0.760337, 0.322744);
\draw[line width=1pt, color=blue](0.760337, 0.322744) -- (0.412923, 0.183845);
\draw[line width=1pt, color=blue](0.412923, 0.183845) -- (0.235640, 0.109881);
\draw[line width=1pt, color=blue](0.235640, 0.109881) -- (0.305590, 0.149046);
\draw[line width=1pt, color=blue](0.305590, 0.149046) -- (0.502528, 0.256051);
\draw[line width=1pt, color=blue](0.502528, 0.256051) -- (0.455601, 0.242247);
\draw[line width=1pt, color=blue](0.455601, 0.242247) -- (0.180172, 0.099871);
\draw[line width=1pt, color=blue](0.180172, 0.099871) -- (0.677232, 0.391000);
\draw[line width=1pt, color=blue](0.677232, 0.391000) -- (0.486871, 0.292542);
\draw[line width=1pt, color=blue](0.486871, 0.292542) -- (0.859921, 0.537338);
\draw[line width=1pt, color=blue](0.859921, 0.537338) -- (0.868863, 0.564246);
\draw[line width=1pt, color=blue](0.868863, 0.564246) -- (0.538874, 0.363475);
\draw[line width=1pt, color=blue](0.538874, 0.363475) -- (0.616002, 0.431329);
\draw[line width=1pt, color=blue](0.616002, 0.431329) -- (0.496736, 0.360900);
\draw[line width=1pt, color=blue](0.496736, 0.360900) -- (0.583004, 0.439325);
\draw[line width=1pt, color=blue](0.583004, 0.439325) -- (0.501175, 0.391561);
\draw[line width=1pt, color=blue](0.501175, 0.391561) -- (0.276664, 0.224038);
\draw[line width=1pt, color=blue](0.276664, 0.224038) -- (0.291097, 0.244259);
\draw[line width=1pt, color=blue](0.291097, 0.244259) -- (0.246035, 0.213875);
\draw[line width=1pt, color=blue](0.246035, 0.213875) -- (0.126335, 0.113752);
\draw[line width=1pt, color=blue](0.126335, 0.113752) -- (0.215018, 0.200508);
\draw[line width=1pt, color=blue](0.215018, 0.200508) -- (0.240259, 0.232016);
\draw[line width=1pt, color=blue](0.240259, 0.232016) -- (0.181019, 0.181019);
\draw[line width=1pt, color=blue](0.181019, 0.181019) -- (0.380673, 0.394198);
\draw[line width=1pt, color=blue](0.380673, 0.394198) -- (0.156860, 0.168211);
\draw[line width=1pt, color=blue](0.156860, 0.168211) -- (0.321183, 0.356710);
\draw[line width=1pt, color=blue](0.321183, 0.356710) -- (0.274233, 0.315469);
\draw[line width=1pt, color=blue](0.274233, 0.315469) -- (0.105417, 0.125631);
\draw[line width=1pt, color=blue](0.105417, 0.125631) -- (0.231590, 0.285990);
\draw[line width=1pt, color=blue](0.231590, 0.285990) -- (0.269660, 0.345149);
\draw[line width=1pt, color=blue](0.269660, 0.345149) -- (0.226282, 0.300287);
\draw[line width=1pt, color=blue](0.226282, 0.300287) -- (0.311526, 0.428779);
\draw[line width=1pt, color=blue](0.311526, 0.428779) -- (0.346440, 0.494768);
\draw[line width=1pt, color=blue](0.346440, 0.494768) -- (0.297491, 0.441048);
\draw[line width=1pt, color=blue](0.297491, 0.441048) -- (0.290837, 0.447850);
\draw[line width=1pt, color=blue](0.290837, 0.447850) -- (0.305233, 0.488476);
\draw[line width=1pt, color=blue](0.305233, 0.488476) -- (0.298722, 0.497157);
\draw[line width=1pt, color=blue](0.298722, 0.497157) -- (0.261000, 0.452065);
\draw[line width=1pt, color=blue](0.261000, 0.452065) -- (0.188106, 0.339352);
\draw[line width=1pt, color=blue](0.188106, 0.339352) -- (0.186850, 0.351413);
\draw[line width=1pt, color=blue](0.186850, 0.351413) -- (0.235167, 0.461541);
\draw[line width=1pt, color=blue](0.235167, 0.461541) -- (0.202527, 0.415243);
\draw[line width=1pt, color=blue](0.202527, 0.415243) -- (0.138619, 0.297269);
\draw[line width=1pt, color=blue](0.138619, 0.297269) -- (0.104938, 0.235695);
\draw[line width=1pt, color=blue](0.104938, 0.235695) -- (0.100809, 0.237490);
\draw[line width=1pt, color=blue](0.100809, 0.237490) -- (0.140103, 0.346767);
\draw[line width=1pt, color=blue](0.140103, 0.346767) -- (0.103210, 0.268871);
\draw[line width=1pt, color=blue](0.103210, 0.268871) -- (0.064300, 0.176662);
\draw[line width=1pt, color=blue](0.064300, 0.176662) -- (0.035812, 0.104007);
\draw[line width=1pt, color=blue](0.035812, 0.104007) -- (0.027812, 0.085595);
\draw[line width=1pt, color=blue](0.027812, 0.085595) -- (0.018403, 0.060192);
\draw[line width=1pt, color=blue](0.018403, 0.060192) -- (0.014800, 0.051613);
\draw[line width=1pt, color=blue](0.014800, 0.051613) -- (0.011745, 0.043831);
\draw[line width=1pt, color=blue](0.011745, 0.043831) -- (0.009181, 0.036822);
\draw[line width=1pt, color=blue](0.009181, 0.036822) -- (0.007054, 0.030556);
\draw[line width=1pt, color=blue](0.007054, 0.030556) -- (0.005314, 0.025002);
\draw[line width=1pt, color=blue](0.005314, 0.025002) -- (0.003912, 0.020126);
\draw[line width=1pt, color=blue](0.003912, 0.020126) -- (0.002802, 0.015894);
\draw[line width=1pt, color=blue](0.002802, 0.015894) -- (0.001943, 0.012267);
\draw[line width=1pt, color=blue](0.001943, 0.012267) -- (0.001294, 0.009207);
\draw[line width=1pt, color=blue](0.001294, 0.009207) -- (0.000819, 0.006673);
\draw[line width=1pt, color=blue](0.000819, 0.006673) -- (0.000486, 0.004624);
\draw[line width=1pt, color=blue](0.000486, 0.004624) -- (0.000264, 0.003016);
\draw[line width=1pt, color=blue](0.000264, 0.003016) -- (0.000126, 0.001804);
\draw[line width=1pt, color=blue](0.000126, 0.001804) -- (0.000049, 0.000944);
\draw[line width=1pt, color=blue](0.000049, 0.000944) -- (0.000014, 0.000387);
\draw[line width=1pt, color=blue](0.000014, 0.000387) -- (0.000002, 0.000089);
\draw[line width=1pt, color=blue](0.000002, 0.000089) -- (0.000000, 0.000000);
\draw[line width=1pt, color=blue](0.000000, 0.000000) -- (-0.000002, 0.000089);
\draw[line width=1pt, color=blue](-0.000002, 0.000089) -- (-0.000014, 0.000387);
\draw[line width=1pt, color=blue](-0.000014, 0.000387) -- (-0.000049, 0.000944);
\draw[line width=1pt, color=blue](-0.000049, 0.000944) -- (-0.000126, 0.001804);
\draw[line width=1pt, color=blue](-0.000126, 0.001804) -- (-0.000264, 0.003016);
\draw[line width=1pt, color=blue](-0.000264, 0.003016) -- (-0.000486, 0.004624);
\draw[line width=1pt, color=blue](-0.000486, 0.004624) -- (-0.000819, 0.006673);
\draw[line width=1pt, color=blue](-0.000819, 0.006673) -- (-0.001294, 0.009207);
\draw[line width=1pt, color=blue](-0.001294, 0.009207) -- (-0.001943, 0.012267);
\draw[line width=1pt, color=blue](-0.001943, 0.012267) -- (-0.002802, 0.015894);
\draw[line width=1pt, color=blue](-0.002802, 0.015894) -- (-0.003912, 0.020126);
\draw[line width=1pt, color=blue](-0.003912, 0.020126) -- (-0.005314, 0.025002);
\draw[line width=1pt, color=blue](-0.005314, 0.025002) -- (-0.007054, 0.030556);
\draw[line width=1pt, color=blue](-0.007054, 0.030556) -- (-0.009181, 0.036822);
\draw[line width=1pt, color=blue](-0.009181, 0.036822) -- (-0.011745, 0.043831);
\draw[line width=1pt, color=blue](-0.011745, 0.043831) -- (-0.014800, 0.051613);
\draw[line width=1pt, color=blue](-0.014800, 0.051613) -- (-0.018403, 0.060192);
\draw[line width=1pt, color=blue](-0.018403, 0.060192) -- (-0.027812, 0.085595);
\draw[line width=1pt, color=blue](-0.027812, 0.085595) -- (-0.035812, 0.104007);
\draw[line width=1pt, color=blue](-0.035812, 0.104007) -- (-0.064300, 0.176662);
\draw[line width=1pt, color=blue](-0.064300, 0.176662) -- (-0.103210, 0.268871);
\draw[line width=1pt, color=blue](-0.103210, 0.268871) -- (-0.140103, 0.346767);
\draw[line width=1pt, color=blue](-0.140103, 0.346767) -- (-0.100809, 0.237490);
\draw[line width=1pt, color=blue](-0.100809, 0.237490) -- (-0.104938, 0.235695);
\draw[line width=1pt, color=blue](-0.104938, 0.235695) -- (-0.138619, 0.297269);
\draw[line width=1pt, color=blue](-0.138619, 0.297269) -- (-0.202527, 0.415243);
\draw[line width=1pt, color=blue](-0.202527, 0.415243) -- (-0.235167, 0.461541);
\draw[line width=1pt, color=blue](-0.235167, 0.461541) -- (-0.186850, 0.351413);
\draw[line width=1pt, color=blue](-0.186850, 0.351413) -- (-0.188106, 0.339352);
\draw[line width=1pt, color=blue](-0.188106, 0.339352) -- (-0.261000, 0.452065);
\draw[line width=1pt, color=blue](-0.261000, 0.452065) -- (-0.298722, 0.497157);
\draw[line width=1pt, color=blue](-0.298722, 0.497157) -- (-0.305233, 0.488476);
\draw[line width=1pt, color=blue](-0.305233, 0.488476) -- (-0.290837, 0.447850);
\draw[line width=1pt, color=blue](-0.290837, 0.447850) -- (-0.297491, 0.441048);
\draw[line width=1pt, color=blue](-0.297491, 0.441048) -- (-0.346440, 0.494768);
\draw[line width=1pt, color=blue](-0.346440, 0.494768) -- (-0.311526, 0.428779);
\draw[line width=1pt, color=blue](-0.311526, 0.428779) -- (-0.226282, 0.300287);
\draw[line width=1pt, color=blue](-0.226282, 0.300287) -- (-0.269660, 0.345149);
\draw[line width=1pt, color=blue](-0.269660, 0.345149) -- (-0.231590, 0.285990);
\draw[line width=1pt, color=blue](-0.231590, 0.285990) -- (-0.105417, 0.125631);
\draw[line width=1pt, color=blue](-0.105417, 0.125631) -- (-0.274233, 0.315469);
\draw[line width=1pt, color=blue](-0.274233, 0.315469) -- (-0.321183, 0.356710);
\draw[line width=1pt, color=blue](-0.321183, 0.356710) -- (-0.156860, 0.168211);
\draw[line width=1pt, color=blue](-0.156860, 0.168211) -- (-0.380673, 0.394198);
\draw[line width=1pt, color=blue](-0.380673, 0.394198) -- (-0.181019, 0.181019);
\draw[line width=1pt, color=blue](-0.181019, 0.181019) -- (-0.240259, 0.232016);
\draw[line width=1pt, color=blue](-0.240259, 0.232016) -- (-0.215018, 0.200508);
\draw[line width=1pt, color=blue](-0.215018, 0.200508) -- (-0.126335, 0.113752);
\draw[line width=1pt, color=blue](-0.126335, 0.113752) -- (-0.246035, 0.213875);
\draw[line width=1pt, color=blue](-0.246035, 0.213875) -- (-0.291097, 0.244259);
\draw[line width=1pt, color=blue](-0.291097, 0.244259) -- (-0.276664, 0.224038);
\draw[line width=1pt, color=blue](-0.276664, 0.224038) -- (-0.501175, 0.391561);
\draw[line width=1pt, color=blue](-0.501175, 0.391561) -- (-0.583004, 0.439325);
\draw[line width=1pt, color=blue](-0.583004, 0.439325) -- (-0.496736, 0.360900);
\draw[line width=1pt, color=blue](-0.496736, 0.360900) -- (-0.616002, 0.431329);
\draw[line width=1pt, color=blue](-0.616002, 0.431329) -- (-0.538874, 0.363475);
\draw[line width=1pt, color=blue](-0.538874, 0.363475) -- (-0.868863, 0.564246);
\draw[line width=1pt, color=blue](-0.868863, 0.564246) -- (-0.859921, 0.537338);
\draw[line width=1pt, color=blue](-0.859921, 0.537338) -- (-0.486871, 0.292542);
\draw[line width=1pt, color=blue](-0.486871, 0.292542) -- (-0.677232, 0.391000);
\draw[line width=1pt, color=blue](-0.677232, 0.391000) -- (-0.180172, 0.099871);
\draw[line width=1pt, color=blue](-0.180172, 0.099871) -- (-0.455601, 0.242247);
\draw[line width=1pt, color=blue](-0.455601, 0.242247) -- (-0.502528, 0.256051);
\draw[line width=1pt, color=blue](-0.502528, 0.256051) -- (-0.305590, 0.149046);
\draw[line width=1pt, color=blue](-0.305590, 0.149046) -- (-0.235640, 0.109881);
\draw[line width=1pt, color=blue](-0.235640, 0.109881) -- (-0.412923, 0.183845);
\draw[line width=1pt, color=blue](-0.412923, 0.183845) -- (-0.760337, 0.322744);
\draw[line width=1pt, color=blue](-0.760337, 0.322744) -- (-0.456174, 0.184306);
\draw[line width=1pt, color=blue](-0.456174, 0.184306) -- (-0.517204, 0.198536);
\draw[line width=1pt, color=blue](-0.517204, 0.198536) -- (-0.654026, 0.238046);
\draw[line width=1pt, color=blue](-0.654026, 0.238046) -- (-0.593786, 0.204457);
\draw[line width=1pt, color=blue](-0.593786, 0.204457) -- (-0.199722, 0.064894);
\draw[line width=1pt, color=blue](-0.199722, 0.064894) -- (-0.700015, 0.214016);
\draw[line width=1pt, color=blue](-0.700015, 0.214016) -- (-0.446025, 0.127896);
\draw[line width=1pt, color=blue](-0.446025, 0.127896) -- (-0.185458, 0.049693);
\draw[line width=1pt, color=blue](-0.185458, 0.049693) -- (-0.424990, 0.105962);
\draw[line width=1pt, color=blue](-0.424990, 0.105962) -- (-0.233849, 0.053988);
\draw[line width=1pt, color=blue](-0.233849, 0.053988) -- (-0.144766, 0.030771);
\draw[line width=1pt, color=blue](-0.144766, 0.030771) -- (-0.261113, 0.050755);
\draw[line width=1pt, color=blue](-0.261113, 0.050755) -- (-0.352561, 0.062166);
\draw[line width=1pt, color=blue](-0.352561, 0.062166) -- (-0.971885, 0.153932);
\draw[line width=1pt, color=blue](-0.971885, 0.153932) -- (-0.566433, 0.079607);
\draw[line width=1pt, color=blue](-0.566433, 0.079607) -- (-0.958800, 0.117726);
\draw[line width=1pt, color=blue](-0.958800, 0.117726) -- (-1.209339, 0.127107);
\draw[line width=1pt, color=blue](-1.209339, 0.127107) -- (-1.444482, 0.126376);
\draw[line width=1pt, color=blue](-1.444482, 0.126376) -- (-1.256931, 0.087893);
\draw[line width=1pt, color=blue](-1.256931, 0.087893) -- (-1.699667, 0.089076);
\draw[line width=1pt, color=blue](-1.699667, 0.089076) -- (-1.666984, 0.058212);
\draw[line width=1pt, color=blue](-1.666984, 0.058212) -- (-1.463777, 0.025550);
\draw[line width=1pt, color=blue](-1.463777, 0.025550) -- (-1.644000, 0.000000);
\draw[line width=1pt, color=blue](-1.644000, 0.000000) -- (-1.891712, -0.033020);
\draw[line width=1pt, color=blue](-1.891712, -0.033020) -- (-1.533066, -0.053536);
\draw[line width=1pt, color=blue](-1.533066, -0.053536) -- (-0.000000, -0.000000);
\draw[line width=1pt, color=blue](-0.000000, -0.000000) -- (-0.000000, -0.000000);
\draw[line width=1pt, color=blue](-0.000000, -0.000000) -- (-0.000000, -0.000000);
    \end{scope}

    \draw [-To,black!30!green,ultra thick](-1.8,-.4) -- (9.32,3.15);
    \draw [-To,black!30!green,ultra thick](13.6,1.3) -- (10.7,2.94);
    \draw [-To,black!30!green,ultra thick](10,3) -- (10,.7);

\end{tikzpicture}

\caption{Illustration of the use of the Murchison Widefield Array (MWA) as a passive radar, showing two transmitters, one satellite and three MWA tiles. The two transmitters' elevation beampatterns (in blue) and the signal path (in green) are also shown.} \label{fig:mwa_radar_illustration}
\end{figure*}

This paper details the use of the MWA as a passive radar for the surveillance of space, illustrated in Figure \ref{fig:mwa_radar_illustration}, and is structured as follows: Section \ref{sec:MWA} {describes} the MWA configuration and its unique aspects as a passive radar receiver. Section \ref{sec:sig_proc} covers the signal processing pipeline and includes the techniques to integrate range-compressed pulses to efficiently match objects in orbit, as well as detailing the complete orbit determination (OD) step in Section \ref{sec:OD}. This OD work builds on earlier work {investigating} uncued detection and initial orbit determination (IOD)~\cite{9559621}. The steps taken to overcome the challenges of using non-cooperative FM radio transmitters (an analogue broadcasting method) amidst the network's unique configuration in Australia are covered in Section \ref{sec:continental_radar}. The paper {includes} some illustrative results in \mbox{Section \ref{sec:resuls}} from a short collection campaign utilising four transmitters around the country at various baseline distances of over 2500~km. The paper concludes with Section \ref{sec:conclusion}.

\section{The Murchison Widefield Array used as a Radar Receiver}\label{sec:MWA}

The MWA is the low-frequency precursor to the SKA, operating in the frequency range of 70--300 MHz.  The main scientific goals of the MWA are to detect radio emissions from neutral hydrogen during the Epoch of Reionisation, to study the Earth's Sun and its heliosphere, Earth's ionosphere, and radio transient phenomena, as well as to map the galactic and extragalactic radio sky~\cite{beardsley2019science}.

The telescope is strategically located in a legislated radio-quiet zone, the Murchison Radio-astronomy Observatory, far from cities and other sources of electromagnetic~interference.

\begin{figure}[ht!]
\begin{center}
\includegraphics[width=0.9\columnwidth]{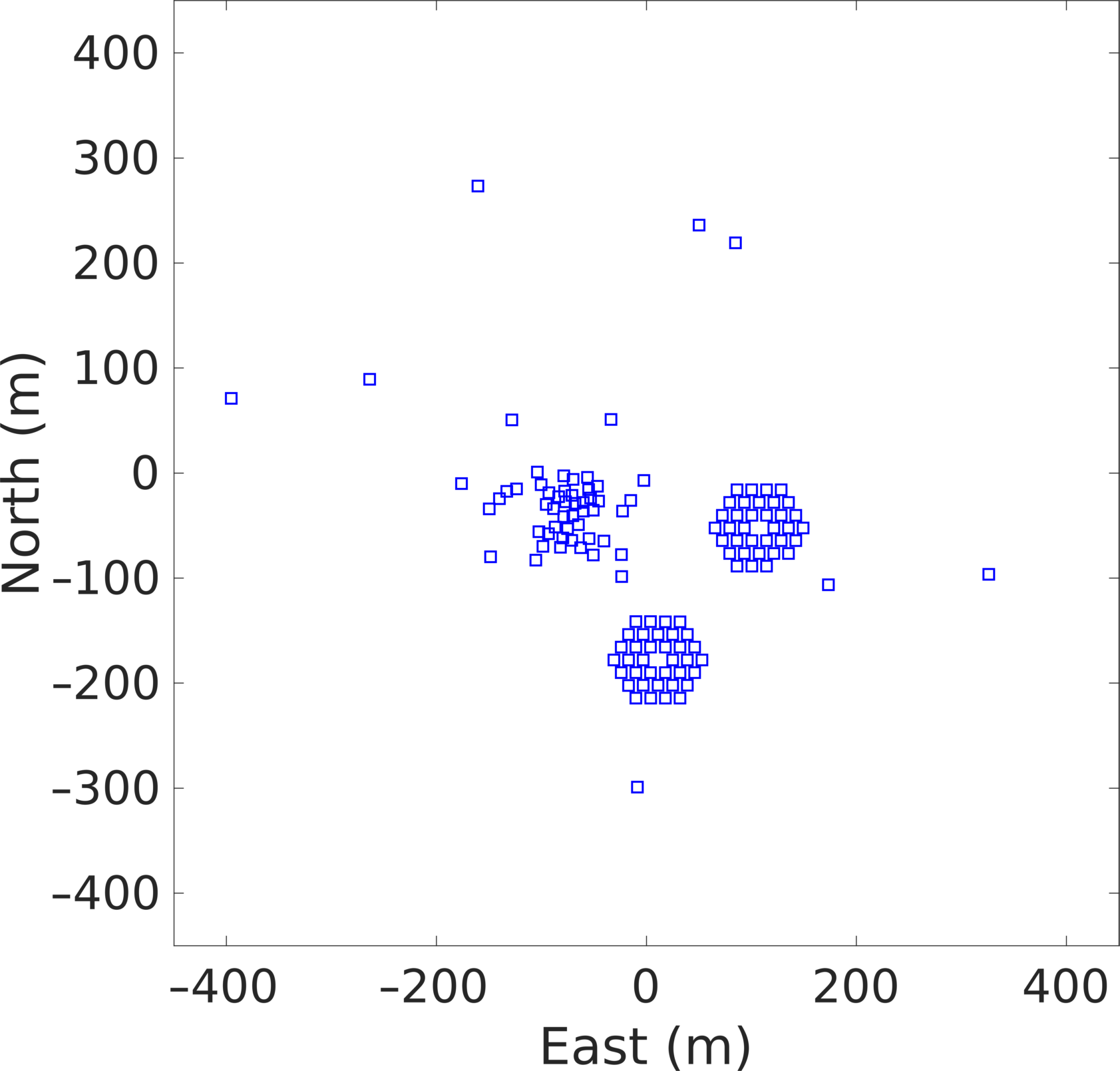}
\end{center}
\caption{A plan view of the Murchison Widefield Array's compact configuration of Phase II, showing the location of the 128 tiles.}
\label{fig:MWA_layout}
\end{figure}

The MWA consists of 4096 dual-polarised wideband dipoles configured into subarrays of 4 $\times$ 4 square grids, referred to as tiles. The grid spacing for each tile is 1.1~m (corresponding to a half-wavelength separation for 136~MHz), and the 16 dipoles are attached to a 5~m $\times$ 5~m steel mesh ground plane. These 256 tiles are distributed over $\sim$15 square kilometres. The current phase of the MWA, Phase II, allows for the use of half of the tiles at any one time, in compact or extended configurations \cite{pase22018article}. Figure \ref{fig:MWA_layout} shows the layout of the compact configuration of the Phase II MWA.

Each tile's 16 antennas are combined with an analogue beamformer to form a tile beam in a particular direction in the sky. The beamwidth is approximately 40$\degree$ at the zenith for FM~frequencies.

The full frequency of each tile is directly sampled at 327.78 MHz covering the array's frequency range of interest, and a polyphase filter bank (PFB) channelises this data to 256 $\times$ 1.28 MHz-wide critically sampled coarse channels. The MWA is able to transfer a user-defined subset of 24 of these 256 channels in real time through to the high time-resolution voltage capture system (HTR-VCS), allowing for an instantaneous bandwidth of 30.72 MHz.

The coarse channels, consisting of 1.28 MHz HTR-VCS data, are then further downsampled to select the various FM frequencies of interest for radar processing.

For typical space surveillance applications, the standard utilisation is a (wide) beam-stare mode, with the tile's analogue beam directed near the zenith.
The net output is 128 channels with a bandwith of (typically) 100 kHz.  These data are then calibrated to remove any impact of the distributed array and phase-align each channel to the antenna. Once calibrated, the data are able to be used for precise electronic beamforming with the distributed array forming a large aperture providing very narrow surveillance beams.
The transmitter reference signals are collected directly at sites close to the transmitter with a small software-defined radio (SDR) system. These SDR reference systems are GPS-disciplined to allow synchronisation with the MWA's surveillance data. Additionally, despite its location in a radio-quiet zone, the MWA is able to detect transmissions from the nearest radio transmitters 300 km away, allowing for the direct observation of nearby reference signals.

\section{Signal Processing}\label{sec:sig_proc}

An issue faced when using a sensor such as the MWA as a space-surveillance radar is the tradeoff between needing to integrate for a longer amount of time for increased sensitivity, and the changing geometry that orbital motion imparts. The precise and narrow beamwidth, resulting from such a large aperture, means that RSOs in LEO will occupy a beam for only a brief instant. For even a short coherent processing interval (CPI), the object in orbit will need to be tracked spatially throughout the CPI. Similarly, the object will need to be tracked in delay and Doppler parameter space as well. This is the fundamental challenge with the detection of RSOs with the MWA; the need to extend the CPI to detect smaller and more distant RSOs is balanced against the difficulty of coherently forming detection signals. This section covers efficient and scalable methods for dealing with this problem by matching the radar-receiver parameters to the orbital motion. This type of track-processing is achievable given the {a priori} information on orbits, whereas for uncued detection and searches, earlier studies have detailed practical methods for forming hypothesised orbits \cite{9559621}. 

Given a large number of potential orbits, either from known tracks, searching a volume of orbits around known tracks, or uncued search hypotheses, the surveillance data detailed in Section \ref{sec:MWA}, and the collected reference signals, are able to be coherently matched to detect an RSO in that orbit.

Rather than directly processing across the received signals, range-compressed pulses are formed for each antenna.  These pulses are then matched to the RSO motion~\cite{1163621,PALMER2011593}. The signals are split into $M$ pulses, each with a duration of $\tau$, such that the CPI length is given by $M\tau$. The pulse length, $\tau$, needs to be sufficiently short to ensure that any change in the Doppler frequency across the pulse is insignificant~\cite{PALMER2011593}. Decreasing the pulse length is equivalent to increasing the maximum unambiguous velocity coverage. For a CPI length of 3~s, $M$ will need to be in the order of 40,000 pulses in order to unambiguously span potential orbital velocities.

Given a reference signal $s_r$ and $N$ tiles that each have a received signal $s_n$ ($n$ ranging from $0$ to $N-1$), the range-compressed pulses are formed by correlation. Given the sample  rate $B$, each pulse consists of $B\tau$ samples, and the pulse compression forming the range-compressed pulse stack is obtained with the following formula:
\begin{IEEEeqnarray}{rCL}
\label{eq:discrete_pulses}
    \chi_n[t,m] =
    \sum_{t'=0}^{B\tau -1}
    s_n[mB\tau+t']{s_r}^*[mB\tau+t'-t]~,
\end{IEEEeqnarray} where $t$ is the fast-time (or delay) sample index, $m$ is the slow-time pulse index, and $t'$ is the fast-time correlation index for the two pulses. {Note that the sample rate, $B$, is treated here as equal to the signal bandwidth, although in practice the true signal bandwidth will vary with the analogue content.}

This pulse stack is formed for each tile, $n$, creating a compressed pulse cuboid. Typically these pulses are coherently integrated simply with a Fourier transform (FT), resolving Doppler. However, for rapidly changing geometries, this will not be sufficient, and instead the phase resulting from the target's motion needs to be matched from pulse to pulse.

Constant radial motion will result in a linear phase rate {across} subsequent pulses, which will be coherently matched by the FT. However, with orbital motion, the target's radial slant range changes rapidly, {as will its Doppler signal}. This results in a complicated signal with the Doppler changing rapidly, which is typically treated as a polynomial phase signal. For orbital motion, matching higher-order terms in this polynomial phase signal results in significant increases in the signal-to-noise ratio (SNR) \cite{4653940,8835821}. The moving object's phase will be different for each antenna as well, resulting in differing polynomial phase signals from each tile which need to be matched in order to be combined.

Although it is possible to form a single beam (and even a moving beam), and then search in Doppler-rate terms (or indeed {vice versa}), the parameter space is far too large and this approach is intractable. Instead, each tile's matching polynomial phase signal is determined from a hypothesised orbit in order to best match the orbital motion. This essentially matches the Doppler phase signal and the spatial phase signal in one process.

\begin{figure}[ht!]

\vspace{-8ex}
\hspace{8ex}\begin{tikzpicture}[dot/.style={draw,fill,circle,inner sep=1pt}]\clip (-2.7,-1) rectangle + (6.2,6);

  \def\a{5.3} 
  \def\b{2} 
  \def\angle{80} 
  \draw [dashed] (0,0) ellipse ({\a} and {\b});
  
  \node[circle,fill,inner sep=1.5pt, label=left:] (O) at (0,0) {};
  \node[circle,fill,inner sep=1.7pt,label=above left:] (X) at (\angle:{{\a}} and {{1.58*\b}}) {};
  \node[circle,fill,inner sep=1.5pt] (RX) at ({\angle-25}:{{\a}} and {{\b}}) {};
  \node[circle,fill,inner sep=1.5pt] (TX) at ({\angle+35}:{{\a}} and {{\b}}) {};
  \draw [thick, ->, shorten >= .09cm] (O) -- (X) node [near end, above left] {$\boldsymbol{r}$};
  \draw [thick, ->, shorten >= .04cm] (O) -- (RX) node [midway, below right] {$\boldsymbol{r}_{rx}$};
  \draw [thick, ->, shorten >= .04cm] (O) -- (TX) node [midway, left] {$\boldsymbol{r}_{tx}$};
  \draw [thick, ->, shorten >= .05cm] (RX) -- (X) node [midway, above right] {$\boldsymbol{\rho}_{rx}$};
  \draw [thick, ->, shorten >= .05cm] (TX) -- (X) node [midway, above left] {$\boldsymbol{\rho}_{tx}$};
  \draw [thick, ->] (X) -- (2.1,3.14) node [above] {$\dot{\boldsymbol{r}}$};
  \node (origin) at (0.4,-.02) {$O$};


  
\end{tikzpicture}
\vspace{-7ex}
\caption{The bistatic radar configuration with the position of an orbital object and its velocity, $\boldsymbol{r}$ and $\dot{\boldsymbol{r}}$, along with the positions of the transmitter and receiver, $\boldsymbol{r}_{rx}$ and $\boldsymbol{r}_{tx}$, as well as the vectors from these sites to the object, $\boldsymbol{\rho}_{rx}$ and $\boldsymbol{\rho}_{tx}.$ The origin, $O$, corresponds to the gravitational centre of the Earth.}

\label{fig:system_geometry}
\end{figure}
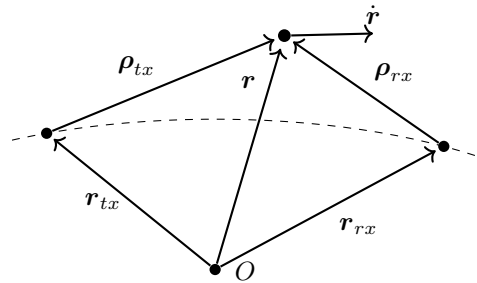

The bistatic radar configuration is illustrated in {Figure} \ref{fig:system_geometry}, with a target at position $\boldsymbol{r}$, relative to the centre of the Earth, with the velocity $\dot{\boldsymbol{r}}$. The receiver's Cartesian location is given by $\boldsymbol{r}_{rx}$ and the transmitter's by $\boldsymbol{r}_{tx}$. The position vectors from the receiver and the transmitter to the RSO are $\boldsymbol{\rho}_{rx}$, and $\boldsymbol{\rho}_{tx}$, respectively. The polynomial phase coefficients are derived from this geometry with an orbital motion model.

The bistatic radar configuration is illustrated in Fig. \ref{fig:system_geometry}, with a target at position $\boldsymbol{r}$, relative to the centre of the Earth, with velocity $\dot{\boldsymbol{r}}$. The receiver's cartesian location is given by $\boldsymbol{r}_{rx}$ and the transmitter's by $\boldsymbol{r}_{tx}$. The position vectors from the receiver and the transmitter to the RSO are $\boldsymbol{\rho}_{rx}$, and $\boldsymbol{\rho}_{tx}$, respectively. The polynomial phase coefficients are derived from this geometry with an orbital motion model.

For the relatively short duration of a single CPI, the motion model assumes that the only force acting on the object in orbit is Earth's gravity.  The acceleration due to the Earth's gravity $\boldsymbol{\ddot{r}}$, is given by the inverse square law:

\begin{equation}
\boldsymbol{\ddot{r}}= -\frac{\mu}{\lvert\boldsymbol{r}\rvert^3}\boldsymbol{r}\label{eq:eci_acceleration},
\end{equation}
where $\mu$ is the standard gravitational parameter for Earth.

This type of orbit, a Keplerian orbit, describes two-body motion and is the simplest orbital model. An orbit is fully determined by the state vector $\boldsymbol{x} =  \begin{bmatrix} \boldsymbol{r} ~ \dot{\boldsymbol{r}} \end{bmatrix}$, and so every future or past motion parameter, and then the phase parameter can be determined from the state vector $\boldsymbol{x}$ at a given time.

Given $\rhovec_{rx} = \boldsymbol{r} - \boldsymbol{r}_{rx}$, $\rhodvec_{rx} = \dot{\boldsymbol{r}} - \dot{\boldsymbol{r}}_{rx}$, etc., the receiver range, range rate, range acceleration and range jerk are given by:
\begin{align}
  \rho_{rx} &= \lvert\rhovec_{rx}\rvert\label{eq:straight_up_slant_range}~,\\
  \label{eq:orbit_doppler}
  \dot{\rho}_{rx} &= \frac{\rhovec_{rx}\cdot\rhodvec_{rx}}{\rho_{rx}}~, \\
  \label{eq:orbit_chirp}
  \ddot{\rho}_{rx} &= -\frac{(\rhovec_{rx}\cdot\rhodvec_{rx})^2}{{\rho_{rx}}^3} +  \frac{|\rhodvec_{rx}|^2 + \rhovec_{rx}\cdot\rhoddvec_{rx}}{\rho_{rx}}~, \\
  \label{eq:orbit_jerk}
  \dddot{\rho}_{rx} &= \begin{multlined}[t]
     3\frac{(\rhovec_{rx}\cdot\rhodvec_{rx})^3}{{\rho_{rx}}^5} \\
     - 3\frac{(\rhovec_{rx}\cdot\rhodvec_{rx})(|\rhodvec_{rx}|^2 + \rhovec_{rx}\cdot\rhoddvec_{rx})}{{\rho_{rx}}^3}\\
     + \frac{3\rhodvec_{rx}\cdot\rhoddvec_{rx} + \rhovec_{rx}\cdot\rhodddvec_{rx}}{\rho_{rx}}~,
  \end{multlined}
\end{align}
and this is similarly the case for the transmitter's terms $\rho_{tx}$, $\dot{\rho_{tx}}$, $\ddot{\rho}_{tx}$ and $\dddot{\rho}_{tx}$. Additionally, $\dddot{\boldsymbol{r}}$ is determined from \eqref{eq:eci_acceleration} and is given by:
\begin{IEEEeqnarray}{rCL}
\dddot{\boldsymbol{r}} =  \frac{3\mu\boldsymbol{r}\cdot\boldsymbol{\dot{r}}}{\lvert\boldsymbol{r}\rvert^5}\boldsymbol{{r}} -\frac{\mu}{\lvert\boldsymbol{r}\rvert^3}\boldsymbol{\dot{r}}~.
\end{IEEEeqnarray}

Note that $\dot{\boldsymbol{r}}_{rx}$, $\ddot{\boldsymbol{r}}_{rx}$ and $\dddot{\boldsymbol{r}}_{rx}$ are the (known) motion terms for the receiver on the Earth's surface, and this is similarly the case for $\dot{\boldsymbol{r}}_{tx}$, $\ddot{\boldsymbol{r}}_{tx}$ and $\dddot{\boldsymbol{r}}_{tx}$ for the transmitter. In this reference frame with a rotating Earth, the MWA is travelling (instantaneously) 47 m/s faster than the southern-most transmitter used in Section \ref{sec:resuls}.

The expressions for the instantaneous bistatic delay and Doppler are
\begin{IEEEeqnarray}{rCL}
\label{eq:ddmap_delay}
    t_D =& &~\frac{1}{c}({\rho}_{rx} + {\rho}_{tx} - \lvert\boldsymbol{r}_{rx} - \boldsymbol{r}_{tx}\rvert)~,\\
\label{eq:ddmap_doppler_shift}
    f_D =& &-\frac{1}{\lambda} (\dot{\rho}_{rx} + \dot{\rho}_{tx})~.
\end{IEEEeqnarray}

The spatial parameters, azimuth and elevation, and their rates, are determined directly from the orbit as well, ensuring that any RSO is spatially tracked throughout the CPI. These parameters are determined from the receiver's slant range vector rotated from an Earth-centred inertial (ECI) geocentric equatorial reference frame to a south-east zenith (SEZ) topocentric-horizon frame. The rotated vector $\boldsymbol{q}$ and its subsequent rates $\dot{\boldsymbol{q}}$ and $\ddot{\boldsymbol{q}}$ are given by the following formula:
\begin{IEEEeqnarray}{rCL}
    \boldsymbol{q} &=& \boldsymbol{D}^{-1} \rhovec_{rx}~,\\
    \dot{\boldsymbol{q}} &=& \boldsymbol{D}^{-1} \rhodvec_{rx}~,\\
    \ddot{\boldsymbol{q}} &=& \boldsymbol{D}^{-1} \rhoddvec_{rx}~,
\end{IEEEeqnarray}
where $\boldsymbol{D}$ is the SEZ to ECI rotation matrix \cite{Vallado2001fundamentals}. {Note in some publications $\boldsymbol{q}$ and $\boldsymbol{\rho}$ are instead written as $\boldsymbol{\rho}_{sez}$ and $\boldsymbol{\rho}_{eci}$ respectively.}

From these topocentric pointings, the azimuth and elevation, $\theta$ and $\phi$, respectively, and their rates, are determined. Given $\boldsymbol{q} = [q_S, q_E, q_Z]^T$, the expressions for the spatial parameters are:

\begin{IEEEeqnarray}{rCL}
    \theta &=& \frac{\pi}{2} - \tan^{-1}\left(\frac{q_E}{q_S}\right)~, \label{eq:azimuth}\\
    \dot{\theta} &=& \frac{\dot{q}_S q_E - \dot{q}_E q_S}{{q_S}^2 + {q_E}^2}~,\\
\ddot{\theta} &=& \frac{1}{{({{q_S}^2 + {q_E}^2})^2}}\big((\ddot{q}_S q_E - \ddot{q}_E q_S)({q_S}^2 + {q_E}^2)\label{eq:azimuth_rate}\\
&& -2(\dot{q}_S q_S + \dot{q}_E q_E)(\dot{q}_S q_E - \dot{q}_E q_S)\big)~,\nonumber\label{eq:azimuth_raterate}
\end{IEEEeqnarray}
and
\begin{IEEEeqnarray}{rCL}
    \phi &=& \tan^{-1}\left(\frac{q_Z}{\sqrt{{q_S}^2 + {q_E}^2}}\right)~,\label{eq:elevation}\\
    \dot{\phi} &=& \frac{\dot{q}_Z - \dot{q}\sin{\phi}}{\sqrt{{q_S}^2 + {q_E}^2}}~,\\
    \ddot{\phi} &=& \frac{1}{{q_S}^2 + {q_E}^2}\bigg( \frac{(\dot{q}\sin{\phi} - \dot{q}_Z  )(\dot{q}_S q_S + \dot{q}_E q_E)}{\sqrt{{q_S}^2 + {q_E}^2}}\label{eq:elevation_rate}\\
    && +(\ddot{q}_Z - \ddot{q}\sin{\phi} - \dot{q}\dot{\phi}\cos{\phi} )\sqrt{{q_S}^2 + {q_E}^2} \bigg)\label{eq:elevation_raterate}~.\nonumber
\end{IEEEeqnarray}

Note that the slant ranges (and their rates) will be unchanged by the rotation, \mbox{$q = \lvert \boldsymbol{q} \rvert = \rho_{rx}$}, and again, this is also the case for the subsequent rates.

A previous study assumed that two terms for each of the spatial parameters were sufficient~\cite{8835821}. However, some particularly fast moving objects, such as rocket bodies in geosynchronous transfer orbits, require additional parameters. The full angular accelerations were required to detect an SL-12 rocket body (NORAD 20082) travelling at 9.6 km/s (in the ECI reference frame).

An equivalent approach is to rotate the tile locations to the ECI frame, in which case the spatial parameters and subsequent beamforming will have a topocentric right ascension and topocentric declination (rather than the azimuth and elevation) \cite{9559621}.

Given these spatial parameters, the polynomial phase coefficients can be determined for both Doppler and spatial aspects. The Doppler phase coefficients for the first four terms are given by their respective-order Taylor series terms:
\begin{IEEEeqnarray}{rCL}
    d_0  = & -\frac{2\pi}{\lambda}(&{\rho}_{rx} + {\rho}_{tx} - \lvert\boldsymbol{r}_{rx} - \boldsymbol{r}_{tx}\rvert)~,\label{eq:doppler_coef_0}\\
    d_1  = & -\frac{2\pi}{\lambda}(&\dot{\rho}_{rx} + \dot{\rho}_{tx})~,\label{eq:doppler_coef_1}\\
    d_2  = & -\frac{~\pi~}{\lambda}(&\ddot{\rho}_{rx} + \ddot{\rho}_{tx})~,\label{eq:doppler_coef_2}\\
    d_3  = & -\frac{\pi}{3\lambda}(&\dddot{\rho}_{rx} + \dddot{\rho}_{tx})~.\label{eq:doppler_coef_3}
\end{IEEEeqnarray}

Similarly, the expression for the spatial coefficients is given by:
\begin{IEEEeqnarray}{rCL}
    b_{n,0}  & =  -\frac{2\pi}{\lambda}(&\boldsymbol{k} \cdot \boldsymbol{u}_n)~,\\
    b_{n,1}  & =  -\frac{2\pi}{\lambda}(&\dot{\boldsymbol{k}} \cdot \boldsymbol{u}_n)~,\\
    b_{n,2}  & =  -\frac{~\pi~}{\lambda}(&\ddot{\boldsymbol{k}} \cdot \boldsymbol{u}_n)~,
\end{IEEEeqnarray}
where $\boldsymbol{u}_n$ is the location of the $n$th tile, and $\boldsymbol{k}$, $\dot{\boldsymbol{k}}$ and $\ddot{\boldsymbol{k}}$ are the wavevector and its rates determined from \eqref{eq:azimuth}--\eqref{eq:elevation_raterate}.

Finally, the resulting matched phase signal for each antenna can be formed by sampling each antenna's polynomial phase signal at time instances $m\tau$, for $m$ $\in$ $[-\frac{M-1}{2}, \dotsc , \frac{M-1}{2}]$:
\begin{IEEEeqnarray}{rCL}
P_n[m]=\me^{-j(b_{n,0} + (b_{n,1} + d_1)m\tau + (b_{n,2} + d_2)(m\tau)^2 + d_3(m\tau)^3 )}.~~~~~
\end{IEEEeqnarray}

This full set of matching phase signals ensures that a potential orbit determined by the state vector $[\boldsymbol{r}$, $\dot{\boldsymbol{r}}]$ will be completely tracked both spatially and in Doppler across every pulse and every tile.

\begin{figure}[ht!]
\begin{center}
\begin{tikzpicture}[scale=.4]]

    \begin{scope}[
            yshift=50,xshift=-300,every node/.append style={            yslant=0.5,xslant=0},yslant=0.5,xslant=0
            ]
        \fill[white,fill opacity=1] (0,0) rectangle (5,5);
        \draw[step=5mm, black] (0,0) grid (5,5);
        \draw[black,very thick] (0,0) rectangle (5,5);
        \fill[red, opacity=0.3] (3,5) rectangle (2.5,-0);
    \end{scope}   
    
    \begin{scope}[
            yshift=30,xshift=-225,every node/.append style={            yslant=0.5,xslant=0},yslant=0.5,xslant=0
            ]
        \fill[white,fill opacity=1] (0,0) rectangle (5,5);
        \draw[step=5mm, black] (0,0) grid (5,5);
        \draw[black,very thick] (0,0) rectangle (5,5);
        \fill[red, opacity=0.3] (3,5) rectangle (2.5,-0);
    \end{scope}   
    
    \begin{scope}[
            yshift=20,xshift=-200,every node/.append style={            yslant=0.5,xslant=0},yslant=0.5,xslant=0
            ]
        \fill[white,fill opacity=1] (0,0) rectangle (5,5);
        \draw[step=5mm, black] (0,0) grid (5,5);
        \draw[black,very thick] (0,0) rectangle (5,5);
        \fill[red, opacity=0.3] (3,5) rectangle (2.5,-0);
    \end{scope}    
    
    \begin{scope}[
            yshift=10,xshift=-175,every node/.append style={            yslant=0.5,xslant=0},yslant=0.5,xslant=0
            ]
        \fill[white,fill opacity=1] (0,0) rectangle (5,5);
        \draw[step=5mm, black] (0,0) grid (5,5);
        \draw[black,very thick] (0,0) rectangle (5,5);
        \fill[red, opacity=0.3] (3,5) rectangle (2.5,-0);
    \end{scope}

\draw[-{Triangle[width=10pt,length=8pt]}, line width=6pt](6.5,7.5) -- (6.5, 5.5);

        \begin{scope}[
            yshift=115,xshift=50]	
    	\node [draw, fill=white] at (0.8,0.1) {$d_1$};
    	\node [draw,fill=green, opacity=0.3]  at (0.8,.1) {$d_1$};
    	
    	\node [draw, fill=white] at (0.8,-1.35) {$d_2$};
    	\node [draw,fill=green, opacity=0.3]  at (0.8,-1.35) {$d_2$};
    	
    	\node [draw, fill=white] at (0.8,-2.8) {$d_3$};
    	\node [draw,fill=green, opacity=0.3]  at (0.8,-2.8) {$d_3$};

    	\node [draw, fill=white] at (3,0.1) {$b_{0,0}$};
    	\node [draw,fill=yellow, opacity=0.3]  at (3,0.1) {$b_{0,0}$};

    	\node [draw, fill=white] at (3,-1.35) {$b_{0,1}$};
    	\node [draw,fill=yellow, opacity=0.3]  at (3,-1.35) {$b_{0,1}$};
    	
    	\node [draw, fill=white] at (3,-2.8) {$b_{0,2}$};
    	\node [draw,fill=yellow, opacity=0.3]  at (3,-2.8) {$b_{0,2}$};

    	\node [draw, fill=white] at (4.9,0.1) {$b_{1,0}$};
    	\node [draw,fill=yellow, opacity=0.3]  at (4.9,0.1) {$b_{1,0}$};

    	\node [draw, fill=white] at (4.9,-1.35) {$b_{1,1}$};
    	\node [draw,fill=yellow, opacity=0.3]  at (4.9,-1.35) {$b_{1,1}$};

    	\node [draw, fill=white] at (4.9,-2.8) {$b_{1,2}$};
    	\node [draw,fill=yellow, opacity=0.3]  at (4.9,-2.8) {$b_{1,2}$};

    	\node [draw, fill=white] at (8.5,0.1) {$b_{N\mbox{-}1,0}$};
    	\node [draw,fill=yellow, opacity=0.3]  at (8.5,0.1) {$b_{N\mbox{-}1,0}$};

        \node[] at (6.4,-1.5) {\normalsize ...};

    	\node [draw, fill=white] at (8.5,-1.35) {$b_{N\mbox{-}1,1}$};
    	\node [draw,fill=yellow, opacity=0.3]  at (8.5,-1.35) {$b_{N\mbox{-}1,1}$};
    
    	\node [draw, fill=white] at (8.5,-2.8) {$b_{N\mbox{-}1,2}$};
    	\node [draw,fill=yellow, opacity=0.3]  at (8.5,-2.8) {$b_{N\mbox{-}1,2}$};

        \node[] at (5,-4.3) {\normalsize phase coefficients};

\end{scope}
    	
    \begin{scope}[
    	yshift=0,xshift=-150,every node/.append style={
    	    yslant=0.5,xslant=0},yslant=0.5,xslant=0
    	             ]
        \fill[white,fill opacity=1] (0,0) rectangle (5,5);
        \draw[black,very thick] (0,0) rectangle (5,5);
        \draw[step=5mm, black] (0,0) grid (5,5) ; 
        \fill[red, opacity=0.3] (3,5) rectangle (2.5,-0);
    \end{scope}

\node[rotate=-20] at (-8.8,1.8) {\Huge{...}};

\draw [decorate,decoration={brace,amplitude=10pt}]
    (-9.7,9.2) -- (-0.7,9.2) node [midway,yshift=20] {\normalsize pulse stack for $N$ antennas};

\node[] at (6,10.5) {\huge \text{[} $\boldsymbol{r}$, $\dot{\boldsymbol{r}}$ \text{]} };

\node[] at (6,8.8) {\normalsize orbit description };

\draw[-{Triangle[width=10pt,length=8pt]}, line width=6pt](-5,-1) -- (-5, -4);
\draw[-{Triangle[width=10pt,length=8pt]}, line width=6pt](6.5,-1) -- (6.5, -4);

    \begin{scope}[
    	yshift=-350,every node/.append
    	             ]             
        \draw[step=8mm, black] (-0.8,0) grid (-8.8,8);
    	\fill[red, opacity=0.3]  (-0.8 ,0) rectangle (-8.8,8);
        \draw[step=8mm, black] (2.39,0) grid (10.4,8);
    	\shade[top color=yellow, bottom color=green, opacity=0.3]  (2.4 ,0) rectangle (3.2,8);
    	\shade[top color=green, bottom color=yellow, opacity=0.3]  (3.2 ,0) rectangle (4,8);
    	
    	\shade[top color=green, bottom color=yellow, opacity=0.3]  (4 ,0) rectangle (4.8,4);
    	\shade[top color=yellow, bottom color=green, opacity=0.3]  (4 ,4) rectangle (4.8,8);
    	
    	\shade[top color=green, bottom color=yellow, opacity=0.3]  (4.8 ,0) rectangle (5.6,2);
    	\shade[top color=yellow, bottom color=green, opacity=0.3]  (4.8 ,2) rectangle (5.6,4);
    	\shade[top color=green, bottom color=yellow, opacity=0.3]  (4.8 ,4) rectangle (5.6,6);
    	\shade[top color=yellow, bottom color=green, opacity=0.3]  (4.8 ,6) rectangle (5.6,8);

    	\shade[top color=green, bottom color=yellow, opacity=0.3]  (5.6 ,0) rectangle (6.4,8);
    	
    	\shade[top color=yellow, bottom color=green, opacity=0.3]  (6.4 ,0) rectangle (7.2,8);
    	
    	\shade[top color=green, bottom color=yellow, opacity=0.3]  (7.2 ,0) rectangle (8.0,8);
    	\shade[top color=yellow, bottom color=green, opacity=0.3]  (8.0 ,0) rectangle (8.8,8);
    	
    	\shade[top color=yellow, bottom color=green, opacity=0.3]  (8.8 ,0) rectangle (9.6,4);
    	\shade[top color=green, bottom color=yellow, opacity=0.3]  (8.8 ,4) rectangle (9.6,8);
    	
    	\shade[top color=green, bottom color=yellow, opacity=0.3]  (9.6 ,0) rectangle (10.4,8);
        
        \draw[-{Triangle[width=12pt,length=6pt]}, line width=4pt](-0.8,3.6) -- (-0.05,3.6);
        \draw[-{Triangle[width=12pt,length=6pt]}, line width=4pt] (2.4,3.6) -- (1.65,3.6);        
        \node[] at (0.8,3.6) {\huge $\bigodot$ };
    \end{scope}
    
    \node[] at (-5 ,-13.3) {\normalsize single range slice };
    \node[] at (6.8,-13.3) {\normalsize phase matching matrix };

    \draw[line width=0.49mm] (0.8,-11.5) circle (0.8) node {\large $\Sigma$};
    \draw[line width=0.49mm] (0.8,-14.5) circle (0.8) node {\normalsize FT};
    
    \draw[-{Triangle[width=12pt,length=6pt]}, line width=4pt](0.8,-9.45) -- (0.8,-10.7);
    
    \draw[-{Triangle[width=12pt,length=6pt]}, line width=4pt](0.8,-12.25) -- (0.8,-13.7);
    
    \draw[-{Triangle[width=12pt,length=6pt]}, line width=4pt](0.8,-15.25) -- (0.8,-16.4);

        \begin{scope}[
    	yshift=-715,xshift=23,every node/.append
    	             ]             
    	             
        \draw[line width=1pt](-7,0) -- (8,0);
        \draw[line width=1pt](8,0) -- (-7,0);
        \draw[ line width=1pt](0,0) -- (0,7);

        \end{scope}
        
        \begin{scope}[
    	yshift=-665,xshift=23,every node/.append
    	]
        \draw[line width=1pt, color=blue](-6.000000, 0.362702) -- (-5.800000,-0.031527);
        \draw[line width=1pt, color=blue](-5.800000, -0.031527) -- (-5.600000,0.357371);
        \draw[line width=1pt, color=blue](-5.600000, 0.357371) -- (-5.400000,-0.102483);
        \draw[line width=1pt, color=blue](-5.400000, -0.102483) -- (-5.200000,-0.062072);
        \draw[line width=1pt, color=blue](-5.200000, -0.062072) -- (-5.000000,0.744849);
        \draw[line width=1pt, color=blue](-5.000000, 0.744849) -- (-4.800000,0.704517);
        \draw[line width=1pt, color=blue](-4.800000, 0.704517) -- (-4.600000,0.708596);
        \draw[line width=1pt, color=blue](-4.600000, 0.708596) -- (-4.400000,0.335749);
        \draw[line width=1pt, color=blue](-4.400000, 0.335749) -- (-4.200000,-0.603743);
        \draw[line width=1pt, color=blue](-4.200000, -0.603743) -- (-4.000000,0.358619);
        \draw[line width=1pt, color=blue](-4.000000, 0.358619) -- (-3.800000,0.815118);
        \draw[line width=1pt, color=blue](-3.800000, 0.815118) -- (-3.600000,0.244447);
        \draw[line width=1pt, color=blue](-3.600000, 0.244447) -- (-3.400000,0.517347);
        \draw[line width=1pt, color=blue](-3.400000, 0.517347) -- (-3.200000,0.363443);
        \draw[line width=1pt, color=blue](-3.200000, 0.363443) -- (-3.000000,-0.151720);
        \draw[line width=1pt, color=blue](-3.000000, -0.151720) -- (-2.800000,0.146936);
        \draw[line width=1pt, color=blue](-2.800000, 0.146936) -- (-2.600000,-0.393641);
        \draw[line width=1pt, color=blue](-2.600000, -0.393641) -- (-2.400000,0.444198);
        \draw[line width=1pt, color=blue](-2.400000, 0.444198) -- (-2.200000,-0.573535);
        \draw[line width=1pt, color=blue](-2.200000, -0.573535) -- (-2.000000,-0.534435);
        \draw[line width=1pt, color=blue](-2.000000, -0.534435) -- (-1.800000,-0.404749);
        \draw[line width=1pt, color=blue](-1.800000, -0.404749) -- (-1.600000,-1.472142);
        \draw[line width=1pt, color=blue](-1.600000, -1.472142) -- (-1.400000,0.719190);
        \draw[line width=1pt, color=blue](-1.400000, 0.719190) -- (-1.200000,0.162595);
        \draw[line width=1pt, color=blue](-1.200000, 0.162595) -- (-1.000000,-0.377464);
        \draw[line width=1pt, color=blue](-1.000000, -0.377464) -- (-0.800000,0.685149);
        \draw[line width=1pt, color=blue](-0.800000, 0.685149) -- (-0.600000,-0.855758);
        \draw[line width=1pt, color=blue](-0.600000, -0.855758) -- (-0.400000,-0.051121);
        \draw[line width=1pt, color=blue](-0.400000, -0.051121) -- (-0.200000,-0.120724);
        \draw[line width=1pt, color=blue](-0.200000, -0.120724) -- (0.000000,0.159603);
        \draw[line width=1pt, color=blue](0.000000, 0.159603) -- (0.200000,0.156429);
        \draw[line width=1pt, color=blue](0.200000, 0.156429) -- (0.400000,-0.432440);
        \draw[line width=1pt, color=blue](0.400000, -0.432440) -- (0.600000,-0.015026);
        \draw[line width=1pt, color=blue](0.600000, -0.015026) -- (0.800000,-0.082440);
        \draw[line width=1pt, color=blue](0.800000, -0.082440) -- (1.000000,0.313854);
        \draw[line width=1pt, color=blue](1.000000, 0.313854) -- (1.200000,0.546633);
        \draw[line width=1pt, color=blue](1.200000, 0.546633) -- (1.400000,0.554637);
        \draw[line width=1pt, color=blue](1.400000, 0.554637) -- (1.600000,-0.431826);
        \draw[line width=1pt, color=blue](1.600000, -0.431826) -- (1.800000,0.038680);
        \draw[line width=1pt, color=blue](1.800000, 0.038680) -- (2.000000,-0.607059);
        \draw[line width=1pt, color=blue](2.000000, -0.607059) -- (2.200000,-0.556750);
        \draw[line width=1pt, color=blue](2.200000, -0.556750) -- (2.400000,-0.003425);
        \draw[line width=1pt, color=blue](2.400000, -0.003425) -- (2.600000,3.966315);
        \draw[line width=1pt, color=blue](2.600000, 3.966315) -- (2.800000,-0.384833);
        \draw[line width=1pt, color=blue](2.800000, -0.384833) -- (3.000000,0.185689);
        \draw[line width=1pt, color=blue](3.000000, 0.185689) -- (3.200000,-0.112792);
        \draw[line width=1pt, color=blue](3.200000, -0.112792) -- (3.400000,0.558678);
        \draw[line width=1pt, color=blue](3.400000, 0.558678) -- (3.600000,-0.544532);
        \draw[line width=1pt, color=blue](3.600000, -0.544532) -- (3.800000,0.016279);
        \draw[line width=1pt, color=blue](3.800000, 0.016279) -- (4.000000,0.276264);
        \draw[line width=1pt, color=blue](4.000000, 0.276264) -- (4.200000,0.550305);
        \draw[line width=1pt, color=blue](4.200000, 0.550305) -- (4.400000,0.772106);
        \draw[line width=1pt, color=blue](4.400000, 0.772106) -- (4.600000,0.042966);
        \draw[line width=1pt, color=blue](4.600000, 0.042966) -- (4.800000,-0.745795);
        \draw[line width=1pt, color=blue](4.800000, -0.745795) -- (5.000000,-0.371151);
        \draw[line width=1pt, color=blue](5.000000, -0.371151) -- (5.200000,-0.530791);
        \draw[line width=1pt, color=blue](5.200000, -0.530791) -- (5.400000,1.175229);
        \draw[line width=1pt, color=blue](5.400000, 1.175229) -- (5.600000,-0.307801);
        \draw[line width=1pt, color=blue](5.600000, -0.307801) -- (5.800000,0.374038);
        \draw[line width=1pt, color=blue](5.800000, 0.374038) -- (6.000000,-0.096209);  
    \node[] at (0,-2.5) {\normalsize Doppler signal };
        \end{scope}

\end{tikzpicture}
\end{center}
\caption{Illustration of the signal processing steps outlined in Section \ref{sec:sig_proc}. An orbital state vector is used to determine the polynomial phase signal coefficients to form a phase-matching matrix. A single range's slow-time signals are matched to the orbit, and combined using this matrix before detection.}
\label{fig:pulse_stack_adjustments}
\end{figure}
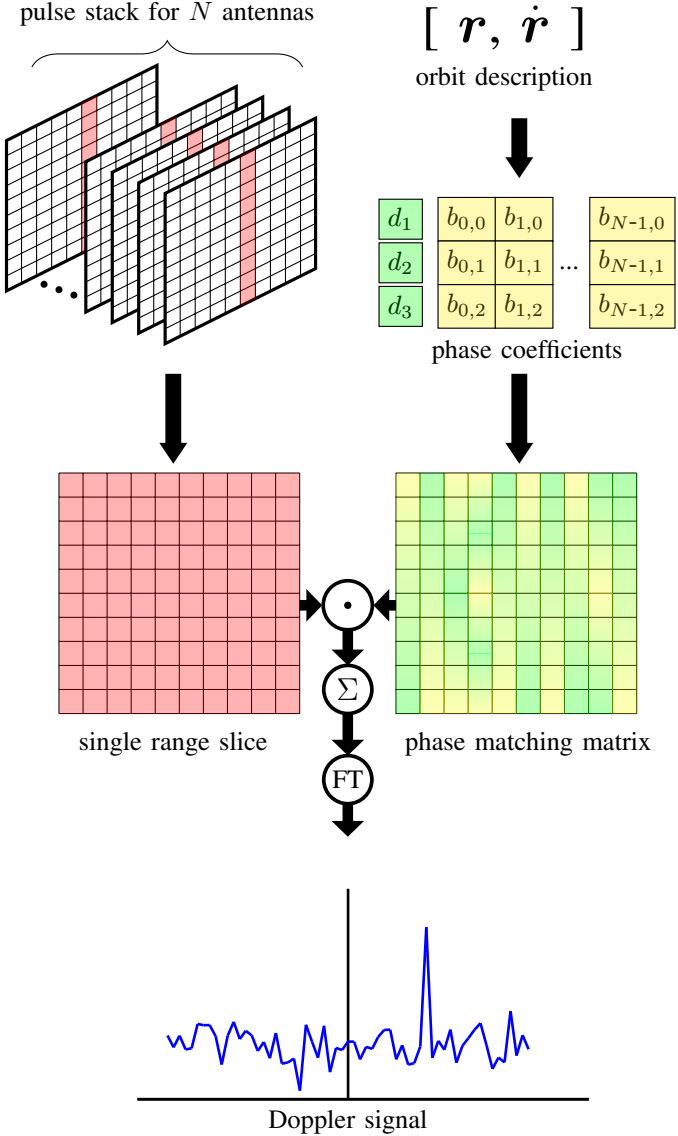 

This phase-matching matrix can then be applied to the data, by applying the polynomial phase signal correction to each tile by forming the Hadamard product between the two. These signals are then coherently combined by summing across each tile to form a single, fully matched, slow time-series for a single range bin, $\chi[m]$. The range bin delay sample is determined by the time delay $t_D$, from \eqref{eq:ddmap_delay}, as well as the sample rate $B$.

\begin{IEEEeqnarray}{rCL}
    \chi[m]  & = & \sum_{n=0}^{N-1} (P_n[m] \odot \chi_n[t,m])\lvert_{t = Bt_D}~. \label{eq:matchy_match}
\end{IEEEeqnarray}

Finally, the Fourier transform will integrate the fully orbital-matched slow-time pulses resolving the slow frequency, or Doppler, signal. This full process is illustrated in {Figure} \ref{fig:pulse_stack_adjustments}. This final signal is passed through a {constant false alarm rate (CFAR)} detector to produce detections of the matched orbit.
    
\begin{IEEEeqnarray}{rCL}
    \chi[f] & = & \left\lvert\sum_{m=-\frac{M-1}{2}}^{\frac{M-1}{2}} \chi[m] \me^{-j2\pi\frac{ f m}{M} }\right\rvert^2~.\label{eq:main_dopp_signal}
\end{IEEEeqnarray}

Note that the $d_0$ term is not included, since the final phase is not of immediate interest; however, the inclusion of $d_1$ ensures any matched target returns will be close to zero Doppler. This allows for a pruned FT implementation. That is, only those frequency bins sufficiently close to zero Doppler need to be determined to cover any potential orbital velocity offset, and also large enough to encompass {sufficient bins allowing for accurate threshold estimation for the CFAR} detector.

The process in \eqref{eq:matchy_match} and {Figure} \ref{fig:pulse_stack_adjustments} only samples a single range bin; creating an entire delay--Doppler map would serve little purpose, since the orbit-derived parameters used to generate that map would only be relevant to a single range. A common pitfall with passive radar and analogue signals is that the ambiguity function is content-dependant; depending on the specific audio signal, the range resolution can be quite poor~\cite{ringer1999waveform}. By processing a single range bin, this issue can be avoided, or at least moderated. Although signal content might result in a poor ambiguity function, by matching orbits directly, the orbit-derived parameters will vary across range bins and only one orbit will have the best-matched~parameters.

By forming detections with orbit-derived parameters, every detection will be associated with an orbital track with some confidence, since the beamforming has followed the orbit through the CPI. This associated trajectory greatly assists ongoing tracking and detection-track association. Additionally, having a {known trajectory estimate is required for the OD step outlined in Section \ref{sec:OD}.}

This process is entirely flexible and the motion model can be extended to more complicated orbital models, such as incorporating an oblate Earth or other perturbing forces. The measurement model can also be tailored, rather than applying far-field beamforming which is suitable for the MWA's compact configuration.  The matched signals are able to be readily extended to near-field beamforming. Instead of calculating beamforming coefficients as well as Doppler coefficients, Doppler coefficients (\eqref{eq:doppler_coef_0}--\eqref{eq:doppler_coef_3}) can be determined for each tile's location and the resulting matched signal can be determined as before.

\section{Orbit Determination}\label{sec:OD}

Given an orbital track, either from an RSO catalogue or an initial orbit hypothesis, the six dimensional positions and velocities are determined and the processing steps described in Section \ref{sec:sig_proc} are applied. The position and velocity state vectors, or indeed the orbital elements, can be adjusted to form search volumes for RSOs, either to detect manoeuvred targets or to update an old track. If an RSO is detected, a series of associated measurements will be produced, although the process utilises the six-dimensional state vectors.  The measurements are produced in the standard radar measurements of azimuth, elevation, bistatic-range, and Doppler.

The number of measurement parameters is extendable in many measurement dimensions.  As covered in previous sections, there is a need to account for higher-order motion parameters such as Doppler rates as well as spatial rates. These dimensions are not searched in the processing steps, as that would result in an intractable search space. Instead, only azimuth, elevation, bistatic range, and Doppler are the adjusted measurement parameters (either directly or via the orbital elements); the first three are searched over as part of the Cartesian location and the latter, Doppler, is searched over via the FT as part of the final Doppler-resolving step. If sufficient computational resources existed, it may be possible to independently search through higher-order parameters, which would allow them to be included as part of the OD step.

Orbit determination is achieved using the batch least-squares method outlined in \cite{montenbruck2012satellite}. This method fits an orbit to a track (or collection) of measurements $\boldsymbol{z}$. The measurements vector $\boldsymbol{z}$ consists of $k$ measurements such that  $\boldsymbol{z} = {\begin{bmatrix} \boldsymbol{z}_0 & \boldsymbol{z}_1 & \dotso & \boldsymbol{z}_{k-1} \end{bmatrix}}^T$, each observed at times $t_0$, $t_1$ ... $t_{k-1}$,  with each measurement consisting of the detected delay, Doppler, azimuth and elevation such that  $\boldsymbol{z}_i = {\begin{bmatrix} {t_D}_i & {f_D}_i & \theta_i & \phi_i \end{bmatrix}}^T$.

If the function $\boldsymbol{f}$ maps a state vector $\boldsymbol{x}$ to its respective measurement parameters at times $t_0$, $t_1$\dotso$t_{k-1}$ (for a single pass, two-body orbit propagation is used such that $\boldsymbol{f}$ consists of  Equations \eqref{eq:ddmap_delay}, \eqref{eq:ddmap_doppler_shift}, \eqref{eq:azimuth} and \eqref{eq:elevation}; for longer-term orbit determination, more complicated models need to be used), the best orbital fit is the state vector which, when propagated, minimises the residuals between the measurements and the predicted measurements:
\begin{IEEEeqnarray}{rCL}
\hat{\boldsymbol{x}} = \mathop{\arg\!\min}\limits_{\boldsymbol{x}}(\lvert\boldsymbol{z} - \boldsymbol{f}(\boldsymbol{x})\rvert^2) \label{eq:od_minimum}
\end{IEEEeqnarray}

As $\boldsymbol{f}$ is highly non-linear, finding a general minima is not trivial and instead a solution is found by linearising all quantities around an initial state vector $\boldsymbol{x}_0$. This initial solution may be provided a priori from a source such as a previous pass or a space catalogue, or instead from the detections directly using an IOD method. The residuals, $\boldsymbol{\epsilon}$, can then be~approximated:
\begin{IEEEeqnarray}{rCL}
\boldsymbol{\epsilon} & = & \boldsymbol{z} - \boldsymbol{f}(\boldsymbol{x})\\
& \approx & \boldsymbol{z} - \boldsymbol{f}(\boldsymbol{x}_0) - \frac{\partial \boldsymbol{f}}{\partial \boldsymbol{x}}(\boldsymbol{x} - \boldsymbol{x}_0) \\
& = & \Delta\boldsymbol{z} - \boldsymbol{F}\Delta\boldsymbol{x}~,
\end{IEEEeqnarray}  with $\Delta\boldsymbol{x} = \boldsymbol{x} - \boldsymbol{x}_0$ being the difference between $\boldsymbol{x}$ and the reference state vector, and $\Delta\boldsymbol{z} = \boldsymbol{z} - \boldsymbol{f}(\boldsymbol{x}_0)$ being the difference between the actual measurements and the predicted measurements for the reference orbit. Additionally, the Jacobian $\boldsymbol{F} = \frac{\partial\boldsymbol{f}(\boldsymbol{x})}{\partial\boldsymbol{x}}\rvert_{\boldsymbol{x} = \boldsymbol{x}_0}$ consists of the partial derivatives of the modelled observations with respect to the state vector.
 
Now, the orbit determination step is achieved by solving a linear least-square problem, with Equation \eqref{eq:od_minimum} simplified as:  
\begin{IEEEeqnarray}{rCL}
{\Delta\boldsymbol{x}}_{ls} = \mathop{\arg\!\min}\limits_{\Delta\boldsymbol{x}}(\lvert\Delta\boldsymbol{z} - \boldsymbol{F}\Delta\boldsymbol{x}\rvert^2) \label{eq:main_linearised_od}~.
\end{IEEEeqnarray}

\begin{figure}[ht!]
\begin{tikzpicture}[dot/.style={draw,fill,circle,inner sep=1pt}]

\node (a) at (0.7,0) {$\boldsymbol{x_0}$};
\node[label=below:{Reference Orbit}] (b) at (7.2,1) {};
\draw[-latex,dashed,bend left=20]  (a) edge (b);
\node (a3) at (0.2,0.4) {${\Delta\boldsymbol{x}}_{ls}$};

\node (a2) at (0.7,0.8) {$\boldsymbol{x}_{ls}$};
\node[label=below:{Estimated Orbit}] (b2) at (7.2,2) {};
\draw[-latex,bend left=20]  (a2) edge (b2);

\draw[-latex]  (a) edge (a2);

\draw (1.4,1) circle (3pt);
\draw (2,1.3) circle (3pt);
\draw (2.6,2.2) circle (3pt);
\draw (3.2,1.5) circle (3pt);
\draw (3.8,2.25) circle (3pt);
\draw (4.4,2.5) circle (3pt);
\draw (5.2,1.8) circle (3pt);
\draw (6,2.6) circle (3pt);

\node (m1) at (3, 2.9) {Measurements};

\end{tikzpicture}
\caption{Illustration of the orbit determination process, starting with a reference orbit and then estimating an orbit {adjustment} to best fit the measurements.}

\label{fig:orbit_determination_process}
\end{figure}
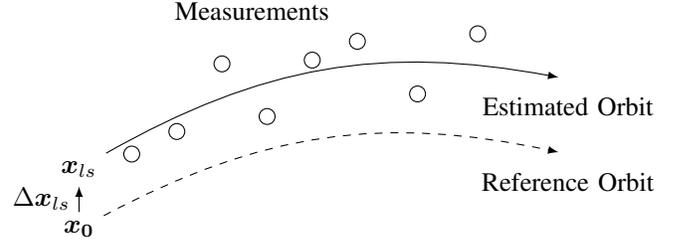

With each detection and track associated with a state vector, $\boldsymbol{x}_0$, a solution to \mbox{Equation~\eqref{eq:main_linearised_od}} is readily determined~\cite{madsen2004methods}. This process is illustrated in Figure \ref{fig:orbit_determination_process}.

In order to compare different measurement types equally, the residuals are normalised by scaling the measurements (and thus, the Jacobians) by the mean measurement error $\sigma_i$. In Equation \eqref{eq:main_linearised_od}, $\boldsymbol{F}$ and $\Delta\boldsymbol{z}$ are replaced by $\boldsymbol{F}\sp{\prime} = \boldsymbol{\Sigma}\boldsymbol{F}$ and $\Delta\boldsymbol{z}\sp{\prime} = \boldsymbol{\Sigma}\Delta\boldsymbol{z}$, with $\boldsymbol{\Sigma}$ being the diagonal matrix $\boldsymbol{\Sigma} = \text{diag}(\sigma_0^{-1}, \sigma_1^{-1}, \dotso,~\sigma_{k-1}^{-1})$.{The mean measurement errors in $\boldsymbol{\Sigma}$ are determined experimentally and also from the measurement resolutions, with the range measurement accuracy determined by the signal bandwidth, the Doppler resolution determined by the CPI length, and the azimuth and elevation resolutions determined by the size of the array aperture. In terms of signal processing, the only aspect that is within the system's control is the Doppler resolution through adjusting the CPI length. It is also possible to further scale the errors using each detection's SNR, such that stronger detections contribute to the orbital fit to a greater extent than the weaker detections \cite{vierinen2017use}.}

With this process, the linearised error covariance matrix is now obtained with the following formula: 
\begin{IEEEeqnarray}{rCL}
    \text{cov}(\boldsymbol{x}_{ls},\boldsymbol{x}_{ls}) = (\boldsymbol{F}^T\boldsymbol{\Sigma}^2\boldsymbol{F})^{-1}~. \label{eq:covariance_lms}
\end{IEEEeqnarray}

The diagonal elements of this covariance matrix yield the {standard deviation of the estimate} of each element of the state vector.

\subsection{Multistatic Orbit Determination}\label{ssec:multistatic_od}

The OD approach is readily extendable to incorporate multistatic returns, with the detections from additional transmitters simply providing extra measurement parameters to fit the orbit. Having each detection associated with a state vector allows for the easy association of multistatic measurements.

A given detection's position is defined by the narrow beamwidth of the electronically steered beam and its intersection with the isorange ellipsoid defined by the time delay from Equation \eqref{eq:ddmap_delay}. However, the range resolution achievable using FM radio signals is quite poor, and although the large aperture allows spatially accurate beams, the volume of the intersection will extend radially. This segment, when intersected with subsequent ellipsoids, will not dramatically improve the estimation, as each subsequent coarse ellipsoid will still be intersected with the identical narrow beam.

Conversely, the target's velocity estimate can be dramatically improved. By expressing~\eqref{eq:ddmap_doppler_shift} in terms of the orbital velocity, $\dot{\boldsymbol{r}}$, it is clear that every Doppler measurement $f_D$ defines a plane of potential velocities:
\begin{IEEEeqnarray}{rCL}
    \left(\frac{\boldsymbol{\rho}_{rx}}{\rho_{rx}} + \frac{\boldsymbol{\rho}_{tx}}{\rho_{tx}}\right)\cdot\boldsymbol{\dot{r}} & = & -\lambda f_D + \frac{\boldsymbol{\rho}_{rx}\cdot\boldsymbol{\dot{r}}_{rx}}{\rho_{rx}} + \frac{\boldsymbol{\rho}_{tx}\cdot\boldsymbol{\dot{r}}_{tx}}{\rho_{tx}}~.~~~~~ \label{eq:doppler_plane} 
\end{IEEEeqnarray}

Additional Doppler returns from different sensors will drastically constrain the extent of possible velocities. Two measurements will define a line, and three or more detections will completely determine (or even overdetermine) the velocity. An accurate velocity estimate is the most important aspect of the orbital state vector estimate, as errors in the velocity estimation will produce increasingly erroneous position estimates when propagated forward. Even if multistatic detections are not coincident, the resulting orbit will be improved for having multiple Doppler measurements to constrain the region of possible~velocities.

Other benefits of multistatic observations include the diversity of coverage, both spatially and in terms of signal content, as well as resilience by making the most of the vast amount of energy being radiated outward. If the bistatic configuration is relying on narrow elevation sidelobes, there will be gaps in illumination. Using multiple transmitters will help ensure any gaps are covered by at least one other sensor. Additionally, analogue FM radio is not necessarily ideal for radar due to the content-dependant nature of the ambiguity function, and diverse options (even multiple stations from the one tower) will provide resilience in the event one station's content is not suitable. FM radio signals have been shown to be well suited for distributed bistatic radar systems \cite{sahr2007lossy}.

\subsection{Initial Orbit Determination}
The process outlined in Section \ref{sec:sig_proc} requires an orbit to match the parameters in order to form detections. Although these orbits can come from a catalogue of tracks, there will still need to be additional work for uncued detection. The six-dimensional search space of all potential orbit state vectors is currently an unreasonably large search space. However, earlier work has shown that the application of radar constraints can greatly simplify the process. By looking for RSOs at their point of minimum bistatic range, and constraining ranges in orbital eccentricity, this search space can be reduced further \cite{9114670}. Akin to creating a spatial fence with receiver beams, this approach only searches a narrow region in orbital parameter space. RSOs that do not pass through the right parameter space on one pass will pass through eventually. Tracked RSOs can then be treated as normal, per the general steps in this section.

There may be thousands of objects from a typical space catalogue observable from the MWA at any one point in time, and matching these tracks (and also searching a region around these tracks) would potentially require matching a hundred thousand orbits. An uncued search of a region, even including a constrained orbital search space, will potentially require matching one million orbits.

{Depending on the available computer memory, a region's hypothesised orbits and resulting parameters can be stored. The phase matrix coefficients, or indeed the full phase matrices, are applicable for all subsequent CPIs and thus, storing them saves on ongoing computation. }

\section{Continental Radar}\label{sec:continental_radar}

\begin{figure}[ht!]
\begin{center}
\includegraphics[width=\columnwidth]{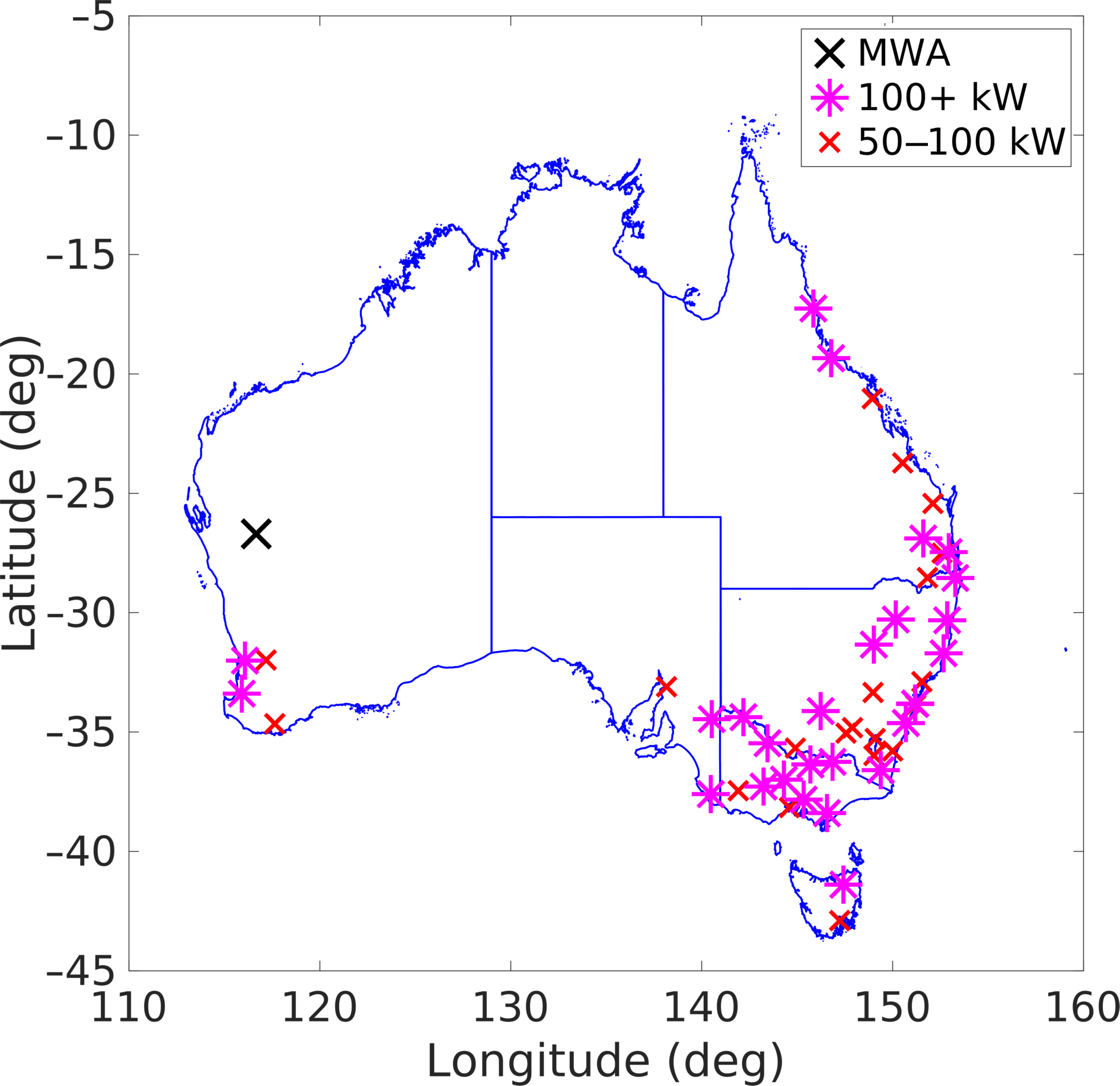}
\end{center}
\caption{A map of all the powerful (greater than 50 kW) FM radio transmitters in Australia.}
\label{fig:Australian_transmitters_power}
\end{figure}

The Australian continent provides a unique context for large-scale FM radio-based passive radar. The majority of the population is concentrated in cities near the coast, particularly in the south-east of the country. The FM radio transmitters are similarly located in the population centres.  This is illustrated by a map of the most powerful FM transmitters in Figure \ref{fig:Australian_transmitters_power}. The MWA is naturally situated far away from these powerful transmitters. This isolation benefits the MWA's astrophysics goals by reducing incident {radio frequency interference (RFI)}. Such a location would ordinarily be less than ideal for terrestrial passive radar purposes, with most passive radar systems designed to be located within the footprint of the illuminating transmitter. However, for detecting satellites at LEO altitudes, such a separation becomes a significant advantage.

\begin{figure}[ht!]
\begin{center}
\includegraphics[width=\columnwidth]{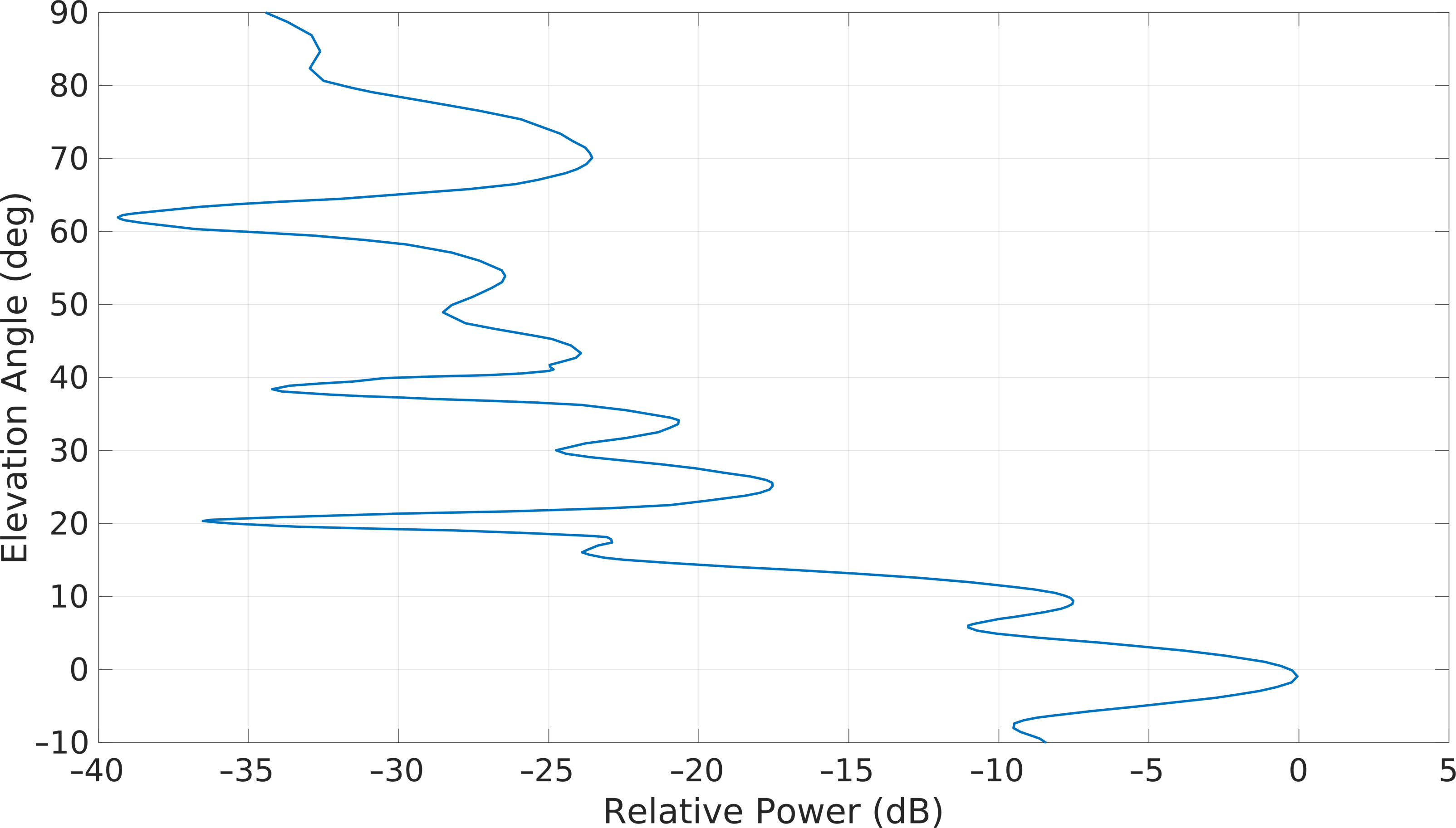}
\end{center}
\caption{Typical FM array's in situ measured antenna pattern via the SixArms airborne radio measurement system~\cite{sixarmsurl}.}
\label{fig:transmit_pattern_el}
\end{figure}

Typical FM radio transmitting antennas consist of six to eight antennas combined to form a beam. These antennas are typically angled (as well as electronically beam-steered) towards the ground in order to direct the maximum amount of energy to the population~\cite{7971960}. This can pose a challenge for satellite illumination, as every effort is made to minimise the amount of wasted energy being radiated outward from Earth, with the main elevation sidelobes being as low as $-$15~dB compared to the main lobe~\cite{7971960}. However, these sidelobes will still provide sufficient illumination, and perhaps some Australian transmitters are not so directive, especially given the widespread and low population density found in some rural areas. Figure \ref{fig:transmit_pattern_el} details a typical FM transmitter pattern.  The main beam is directed to the ground; however, sidelobe levels are not insignificant at elevations of up to 15$\degree$.  At higher elevations, considerable energy is still being radiated in some sidelobes. The patterns of other transmitters may not be so directional.

\begin{figure}[ht!]
\begin{center}
\includegraphics[width=\columnwidth]{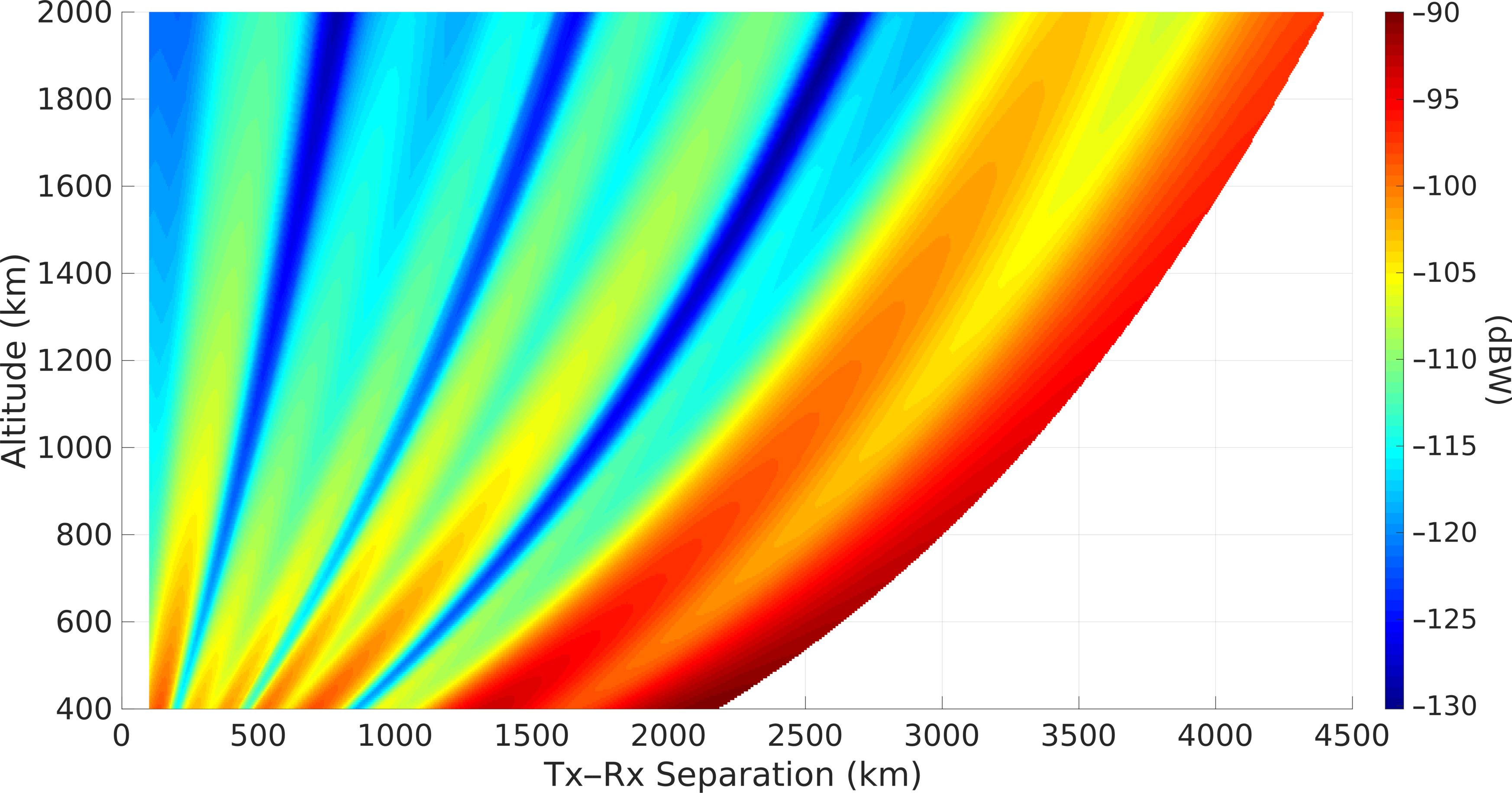}
\end{center}
\caption{Incident power on an object above a receiver, for an equivalent isotropically radiated power (EIRP) of 100~kW, with a beam pattern as shown in Figure \ref{fig:transmit_pattern_el}. Note that this figure is based on a spherical Earth model.}
\label{fig:power_on_orbit}
\end{figure}

Instead of relying on sufficient sidelobes for target illumination, it is possible to make use of the main lobe for illumination. Given the number of transmitters in the south-east of Australia, and the lack of large interfering transmitters in between that region and the MWA, a target above the MWA will be illuminated by the main beam of potentially dozens of significant sources. This is illustrated in {Figure} \ref{fig:power_on_orbit}, showing the incident power on an RSO at various altitudes above a receiver with a wide range of potential transmitter--receiver separation distances. It shows that the loss incurred by the increased transmitter to target distance, $\rho_{tx}$, is more than offset by the transmitter gain of the main lobe, from {Figure} \ref{fig:transmit_pattern_el}. Indeed, the MWA has been used to detect FM radio returns from the moon, undoubtedly from transmitters half-way around the world~\cite{mckinley2012low}.


\begin{figure}[ht!]
\begin{center}
\includegraphics[width=\columnwidth]{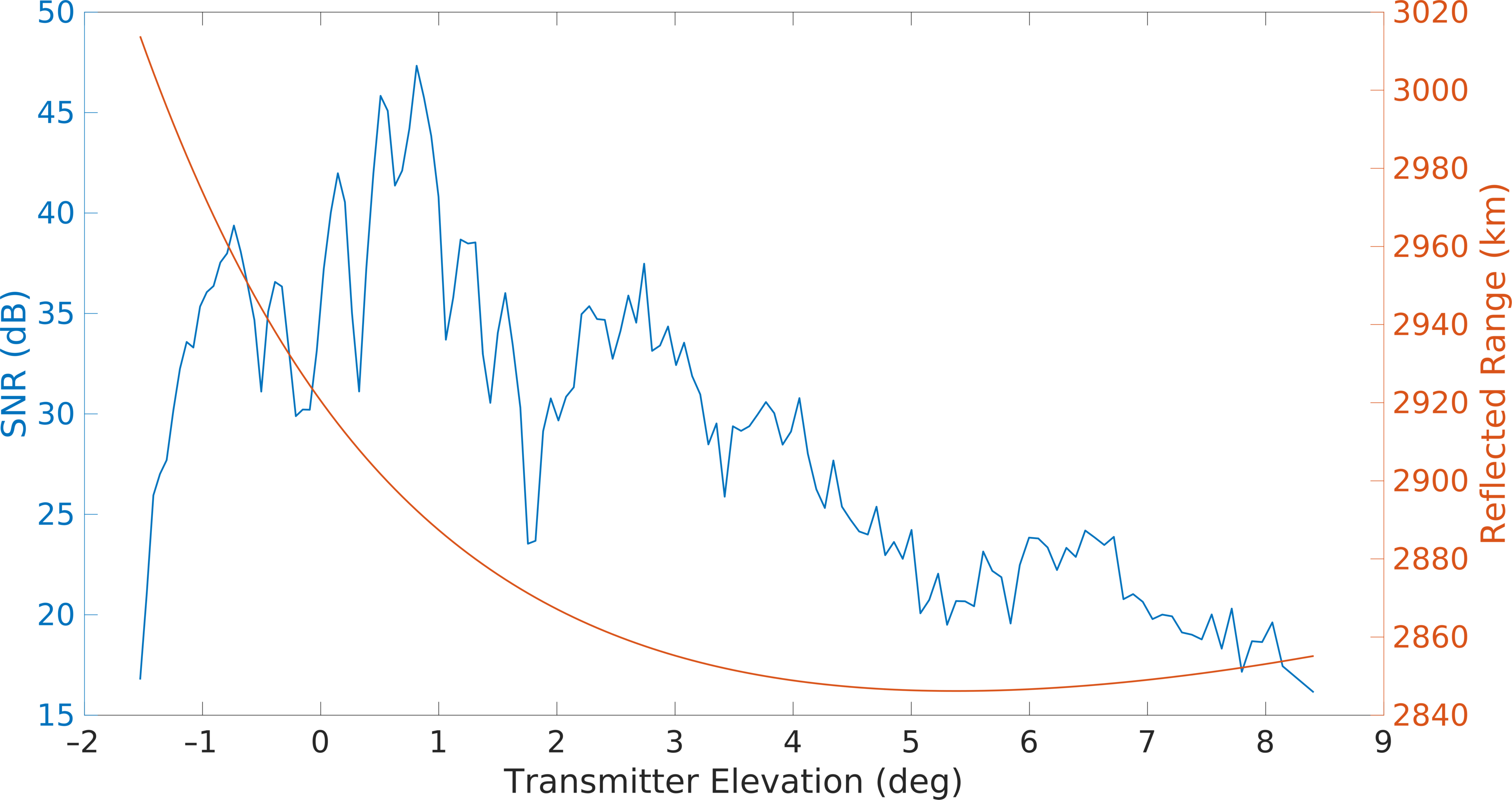}
\end{center}
\caption{An example pass showing the detections of the International Space Station utilising a distant transmitter versus the elevation of the target. The blue line shows the detected signal to noise ratio and the red line shows the signal path length and the reflected range. The reflected range is given by $\rho_{rx} + \rho_{tx}$ and does not include the baseline length to give the full bistatic range, as in Equation \eqref{eq:ddmap_delay}.}
\label{fig:MtGambierElevation}
\end{figure}

The use of the main lobe is illustrated in {Figure} \ref{fig:MtGambierElevation}, showing the SNR of the International Space Station (ISS), utilising a transmitter in Mount Gambier, over 2500~km away from the MWA. This shows that the strongest detections occur at the lowest transmitter elevation angles (despite the larger signal path loss); it even highlights the diffraction of signals along the Earth's surface with detections occurring at transmitter elevations of almost 2$\degree$ below the~horizon.

Finally, the majority of FM transmitters transmit a vertically polarised signal, so the direction relative to the MWA will determine the best receiver polarisation. The vertical polarisation to the south will be coplanar with the MWA's north--south polarisation, and there is a similar situation for transmitters to the east and the MWA's east--west polarisation. However, impacts from effects such as Faraday rotation due to the ionosphere and other factors mean that the presence of matching transmitter and receiver polarisations is not necessarily a decisive factor in the detected SNR.

\section{Results}\label{sec:resuls}

\begin{figure}[ht!]
\begin{center}
\includegraphics[width=\columnwidth]{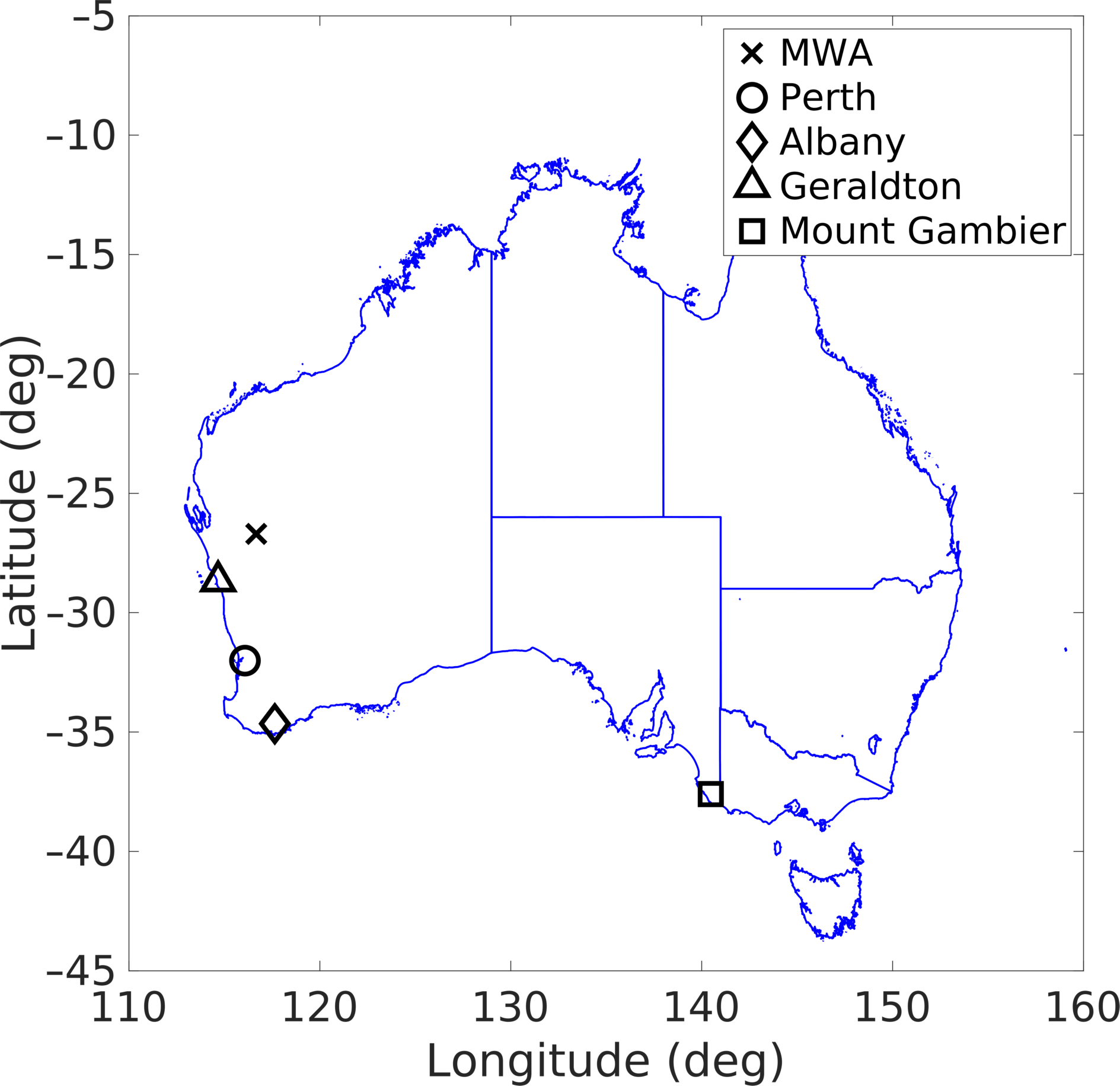}
\end{center}
\caption{A map showing the Murchison Widefield Array as well as the transmitters used to generate the results in \mbox{Section \ref{sec:resuls}}. Details of the transmitters are given in Table \ref{tab:transmitters}.}
\label{fig:placeholder5}
\end{figure}

The results in this section are obtained from 20 min of data consisting of a series of five minute dwells collected in December 2019. {As detailed in Section \ref{sec:MWA}, these dwells were recorded and channelised in real-time and transferred to storage in Perth. Subchannels were selected such that the full national FM band of approximately 20 MHz was collected.} The MWA's analogue beamformer was directed to point at the zenith. In addition to MWA observations, several transmitter reference signals were collected from around the country. These are outlined in Table \ref{tab:transmitters} and shown in {Figure} \ref{fig:placeholder5}. Despite being located in a radio-quiet observatory, a sufficient signal is able to be observed from the nearby transmitters in Geraldton, some 300 km to the south-west of the MWA. Reference observations were collected in Perth {utilising receiver hardware identical to that used in the MWA.} Located at the Curtin Institute of Radio Astronomy {(CIRA), this MWA-like reference receiver is synchronised with the MWA via GPS}. 

Transmitter reference signals were also collected at locations near Albany and Mount Gambier with a SDR setup. {These remote SDR nodes are all GPS-disciplined in order to maintain synchronisation, the collections were manually triggered, and each node was able to record a reference with a bandwidth of 10 MHz. Although 10 MHz is insufficient to collect the full FM band, it is generally sufficient to collect every high-powered FM station from a single site.} The transmitter near Mount Gambier, over 2500~km away from the MWA and situated in the south-east of South Australia, is one of the highest power radio transmitters in the country. It should be noted that for each transmitter there are many different FM radio stations, all potentially being broadcast at different power levels. The figures in Table \ref{tab:transmitters} are simply the maximum licensed power level from that tower, and the true levels may, in fact, be lower.

{Data from these remote SDR devices were then transferred to servers at CIRA in Perth, alongside the MWA-collected data, allowing offline space surveillance processing. This is achieved by downsampling the MWA data, as well as the SDR transmitter reference data, to narrowband signals (typically 100 kHz) matching known FM stations, and then undertaking radar processing as detailed in Section \ref{sec:sig_proc}, utilising a 3~s CPI.}

\vspace{-10pt}

\begin{table}
\begin{center}
\caption{Transmitters utilised in this section's results.}\label{tab:transmitters}
 \begin{tabular}{|c c c|} 
 \hline
 Name (locale) & Maximum Power & Distance from MWA \\ [0.2ex] 
 \hline\hline
 Geraldton  & 30 kW & 295 km  \\ 
 \hline
 Perth  & 100 kW & 591 km  \\
 \hline
 Albany & 50 kW & 886 km  \\
 \hline
 Mount Gambier & 240 kW & 2,524 km  \\
 \hline
 \hline
\end{tabular}
\end{center}
\end{table}

\begin{figure*}[ht!]
\begin{center}
\includegraphics[width=\textwidth]{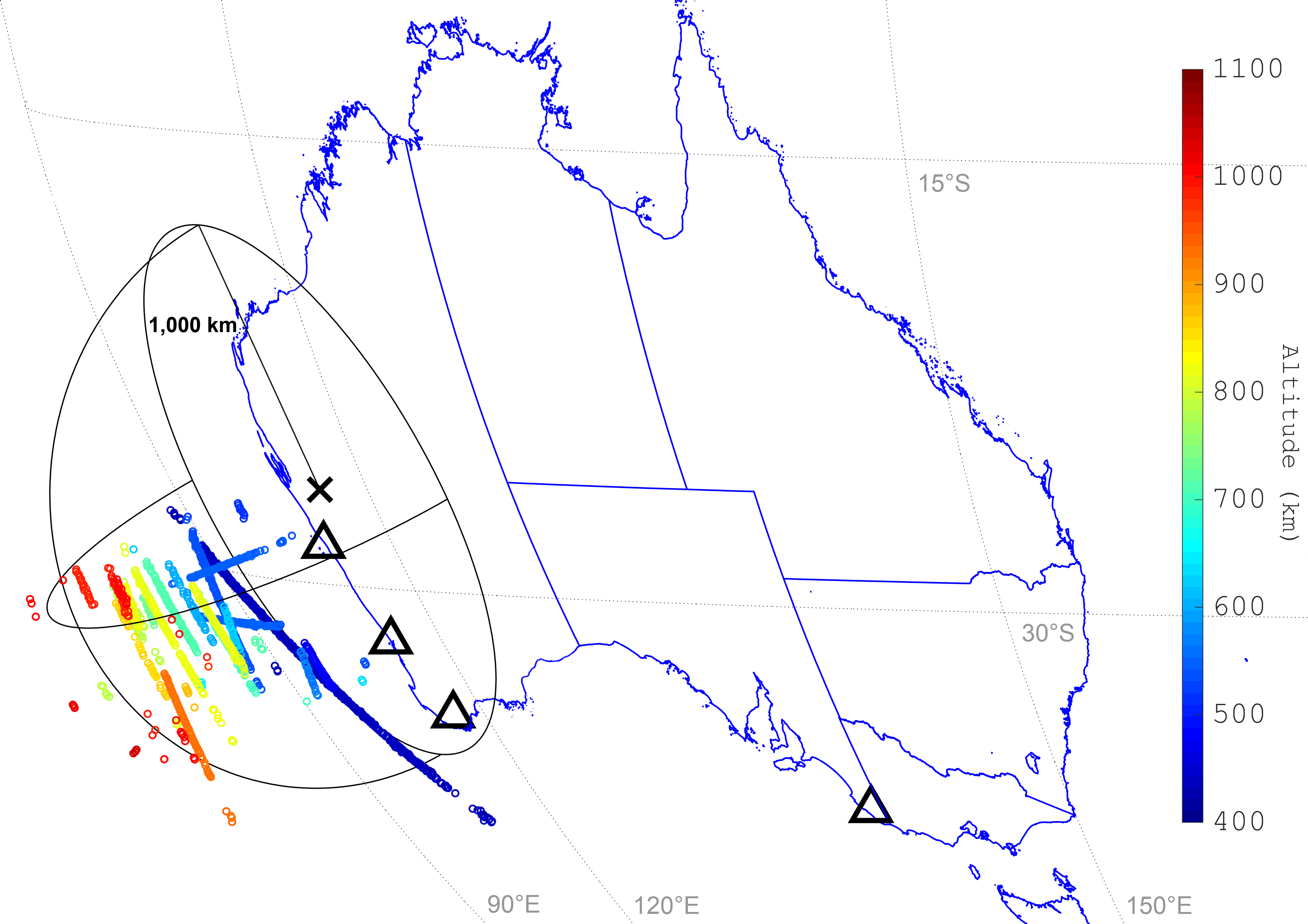}
\end{center}
\caption{Example results showing dozens of tracks' detections above the Murchison Widefield Array (MWA), {with each detection's colour corresponding to its altitude}. {The location of the MWA is shown, denoted by an X. The transmitters are also shown, denoted by the black triangles. Additionally, rings denoting a 1000-km range (from the MWA) are included.}}
\label{fig:placeholder2}
\end{figure*}

\begin{figure*}[ht!]
\begin{center}
\includegraphics[width=\textwidth]{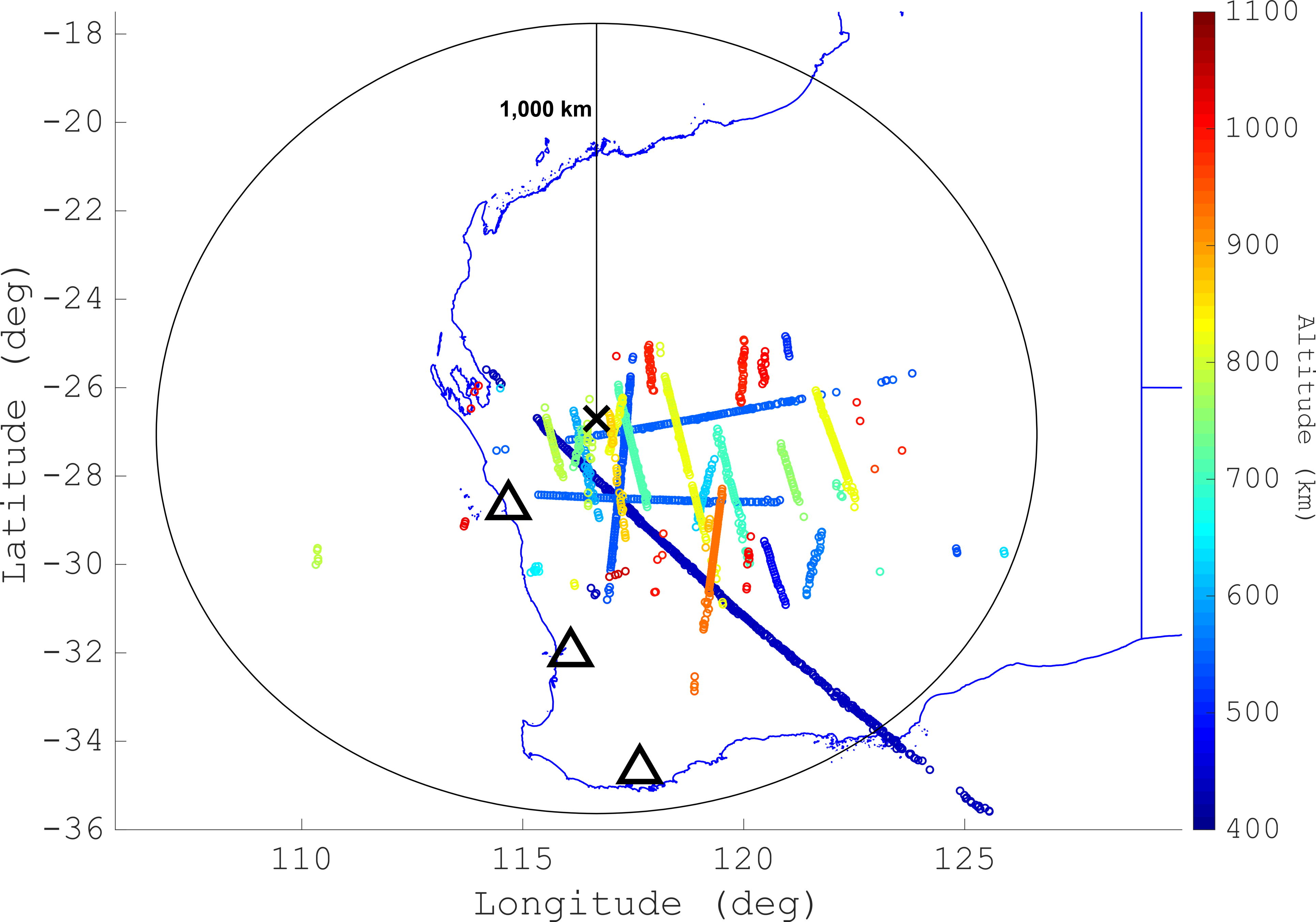}
\end{center}
\caption{Example results,{matching Figure \ref{fig:placeholder2},} showing dozens of tracks' detections above the Murchison Widefield Array (MWA) {with each detection's colour corresponding to its altitude}. {The location of the MWA is shown, denoted by an X. The transmitters are also shown, denoted by the black triangles. Additionally, a ring denoting a 1000-km range (from the MWA) is included.}}
\label{fig:placeholder21}
\end{figure*}

Figures \ref{fig:placeholder2} and \ref{fig:placeholder21} illustrate some of the aggregate detections from these observations. {These detections are formed by parsing the Doppler signal data (from Equation \eqref{eq:main_dopp_signal}) through a cell-averaging CFAR detector. In passive radar processing (and indeed, noise radar), target signals exist against a noise/clutter {pedestal} floor formed by the cross correlation of the reference signal against other unwanted/mismatched signals. The CFAR detector estimates this floor from a local threshold region around the cell that is being {evaluated}, and the SNR is determined by the peak signal against this floor. For these results, we have used a very conservative threshold of 16 dB, greatly minimising the presence of any false detections. System performance would be improved with a more realistic threshold; however, care would need to be taken to ensure false detections are not incorporated into orbital estimates.}

The MWA is able to maintain tracks of many targets at various ranges. During these short and targeted dwells, the MWA was able to detect every large RSO that was within the MWA's main beam at a range of less than 1000~km. The USSPACECOM catalogue defines a large object as having a median radar cross section (RCS) of 1~m$^2$ or greater and a medium object having a median RCS between 0.1~m$^2$ and 1~m$^2$; however, these values are for microwave frequencies which differ from those used in this paper, and should only be taken as a general indicator of size~\cite{USSPACECOM}. Additionally, many detections are found outside these limits, including the detection of medium RSOs, RSOs at longer ranges, and indeed RSOs well outside of the main receiver beam. This is consistent with the earlier theorised performance~\cite{2013AJ....146..103T}. Predictions of the large RSOs with a closest approach of less than 1000~km indicate the MWA would detect over 1800 RSO passes per day, when used in a beam-stare mode pointing at the zenith. However, detections outside these conservative limits, as well as the ability to rapidly adjust the analogue beamforming, suggest the true number will be larger. 

\begin{figure}[ht!]
\begin{center}
\includegraphics[width=\columnwidth]{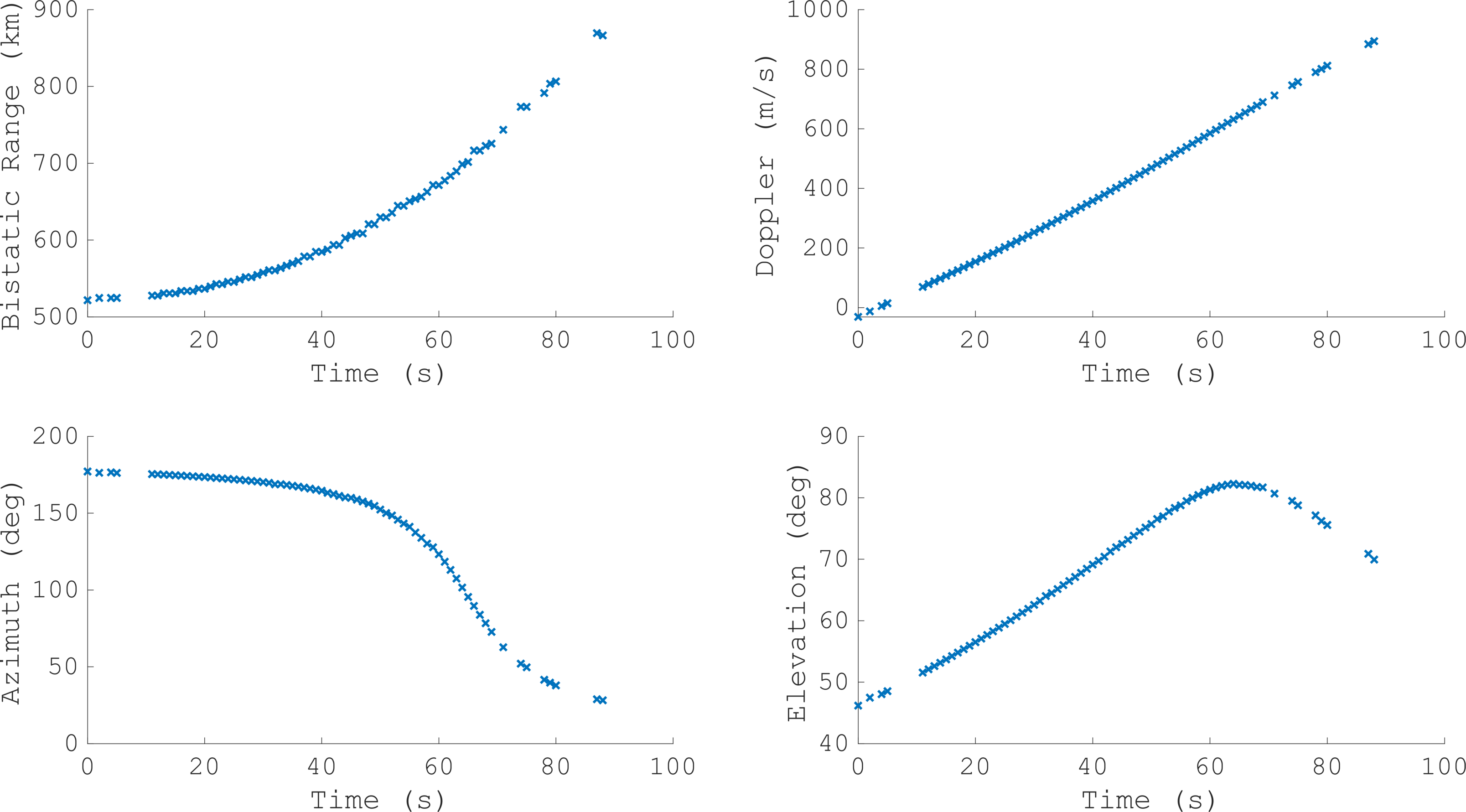}
\end{center}
\caption{The four measurement parameters from the detections of an outbound COSMOS 1707 detected using an FM transmitter in Albany.}
\label{fig:OD_summary_COSMOS_1707_albany_measurements}
\end{figure}

\begin{figure}[ht!]
\begin{center}
\includegraphics[width=\columnwidth]{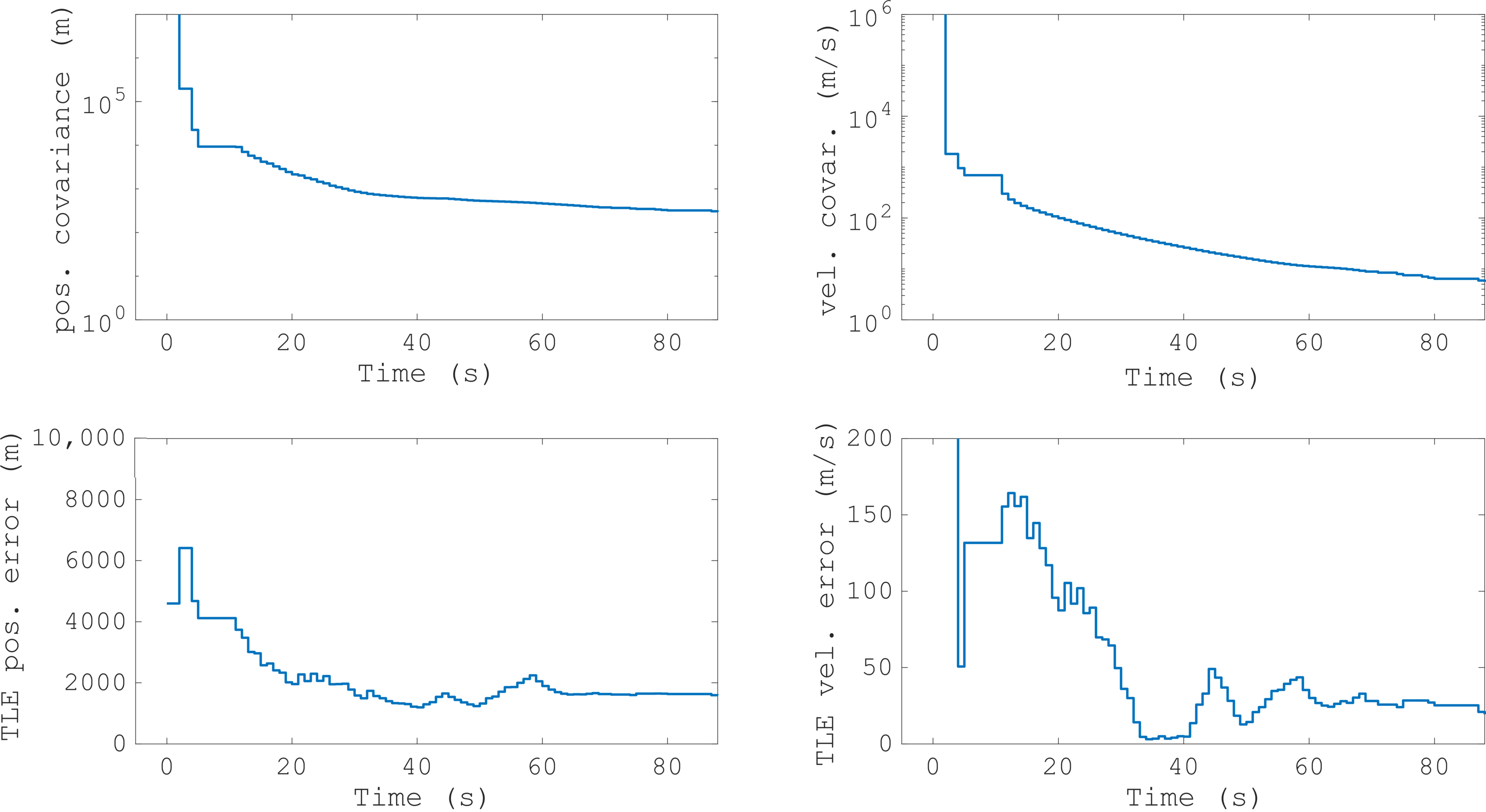}
\end{center}
\caption{The resulting orbit predictions from the measurements from {Figure} \ref{fig:OD_summary_COSMOS_1707_albany_measurements}. The top row shows the covariance of the position estimate and the velocity estimate; the bottom row shows the mean error when compared with the TLE.}
\label{fig:OD_summary_COSMOS_1707_albany_orbit_prop}
\end{figure}

Figure \ref{fig:OD_summary_COSMOS_1707_albany_measurements} shows the detections of an outbound pass of COSMOS 1707 (NORAD 16326), a large (now defunct) satellite. The detections were formed utilising the transmitter near Albany and show the bistatic range, bistatic Doppler, azimuth and elevation. Tracked for almost 90 s, the RSO passes the closest bistatic approach (at zero Doppler) and moves north to the closest approach to the receiver (at its maximum elevation).  Figure \ref{fig:OD_summary_COSMOS_1707_albany_orbit_prop} shows the accuracy of the orbit generated from the COSMOS 1707 measurements. The top row shows the accuracy of the positional and the velocity covariance, from \eqref{eq:covariance_lms}. The bottom row shows the accuracy of the position and velocity estimate in comparison to the two-line element (TLE) ephemeris. The two rows are in general agreement as to the resulting accuracy and the results are significantly improved when compared to the initial study \cite{7944483}. These results are typical of most of the objects the MWA detects with a bistatic configuration.

\begin{figure}[ht!]
\begin{center}
\includegraphics[width=\columnwidth]{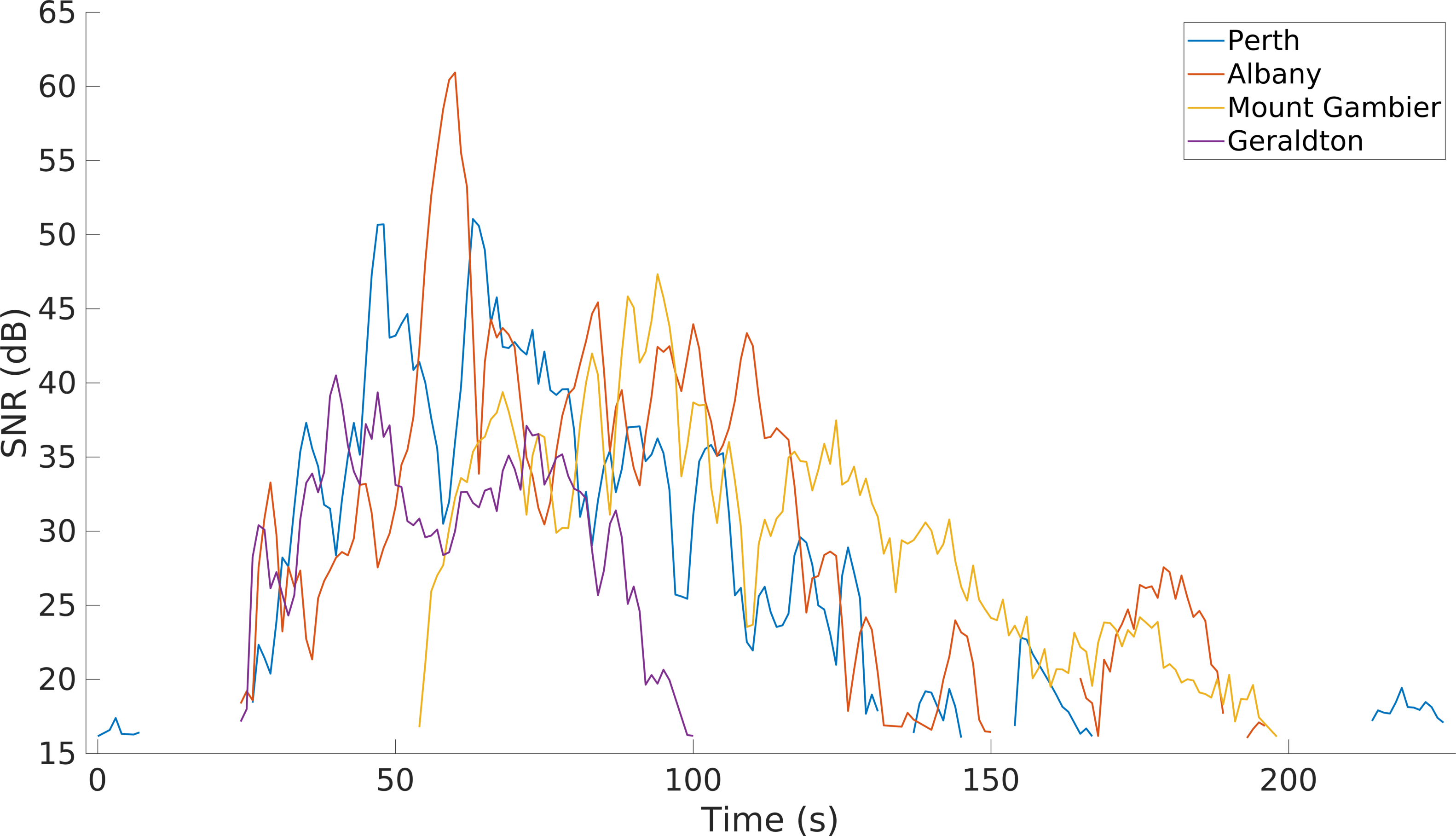}
\end{center}
\caption{An example pass showing the signal to noise ratio (SNR) of International Space Station detections utilising all the transmitters covered in this section.}
\label{fig:placeholder4}
\end{figure}

With many more transmitters at the radar's disposal, there is a scope for increased coverage. Figure \ref{fig:placeholder4} shows the SNR of a pass of the International Space Station for every transmitter collected in this campaign. It shows three minutes of detections with almost 50 s of complete overlap for each bistatic pair. The SNR fluctuations shown (for all transmitters) highlight the variable nature of the illuminator coverage due to changes in the transmitter beampattern as well as variations in bistatic RCS. There may be additional contributing factors such as Faraday rotation. The ISS is detected well outside of the MWA's receiver beam, to an elevation of as low as 5$\degree$ above the horizon.

Although only the ISS and the Hubble Space Telescope were large enough to be detected simultaneously using all transmitters, approximately three quarters of all the detected targets had associated detections from another transmitter. Additionally, every transmitter was able to detect objects that were not detected by any other transmitter, including the comparatively weaker Geraldton site. This highlights that FM broadcast transmissions do not uniformly cover the volume above the MWA, and results will improve with more transmitters being utilised. Despite only being licensed to transmit up to a maximum of 50~kW, the particular configuration of the Albany transmitter, and its elevation sidelobes, produced the largest number of detections of all the transmitters listed in Table \ref{tab:transmitters}.

\begin{figure}[ht!]
\begin{center}
\includegraphics[width=\columnwidth]{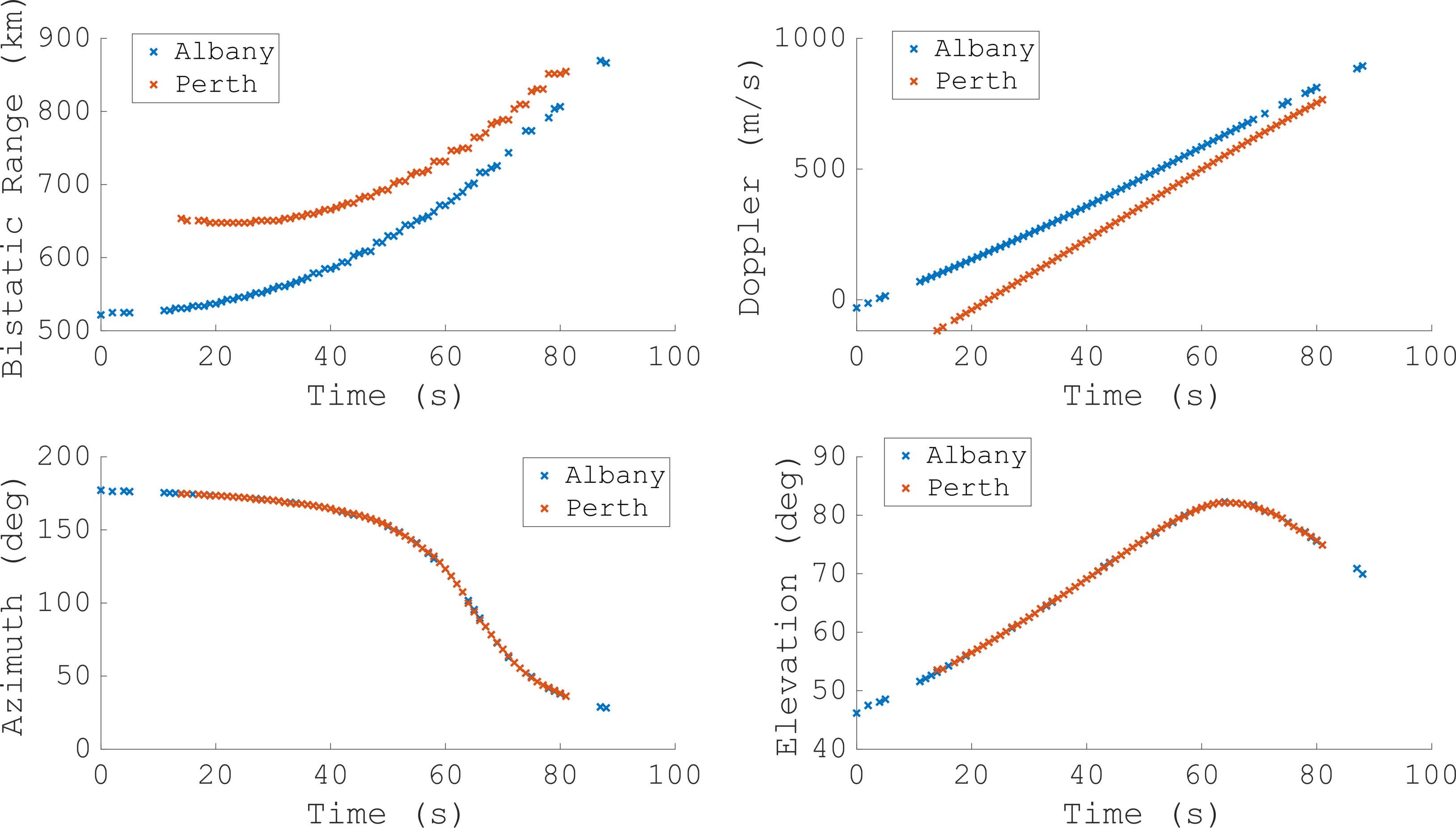}
\end{center}
\caption{The four measurement parameters from the detections of COSMOS 1707 detected using FM transmitters in both Albany and Perth.}
\label{fig:OD_summary_COSMOS_1707_COMBINED_measurements}
\end{figure}

Figure \ref{fig:OD_summary_COSMOS_1707_COMBINED_measurements} shows the detection's measurements for the same RSO pass as in \mbox{Figure \ref{fig:OD_summary_COSMOS_1707_albany_measurements}}; however, this time, the detections from the Perth transmitter are included. The spatial parameters are near-identical as expected, with the differing geometry resulting in differing delay and Doppler tracks. When these are combined together in the OD stage, the results are significantly improved. 

\begin{figure}[ht!]
\begin{center}
\includegraphics[width=\columnwidth]{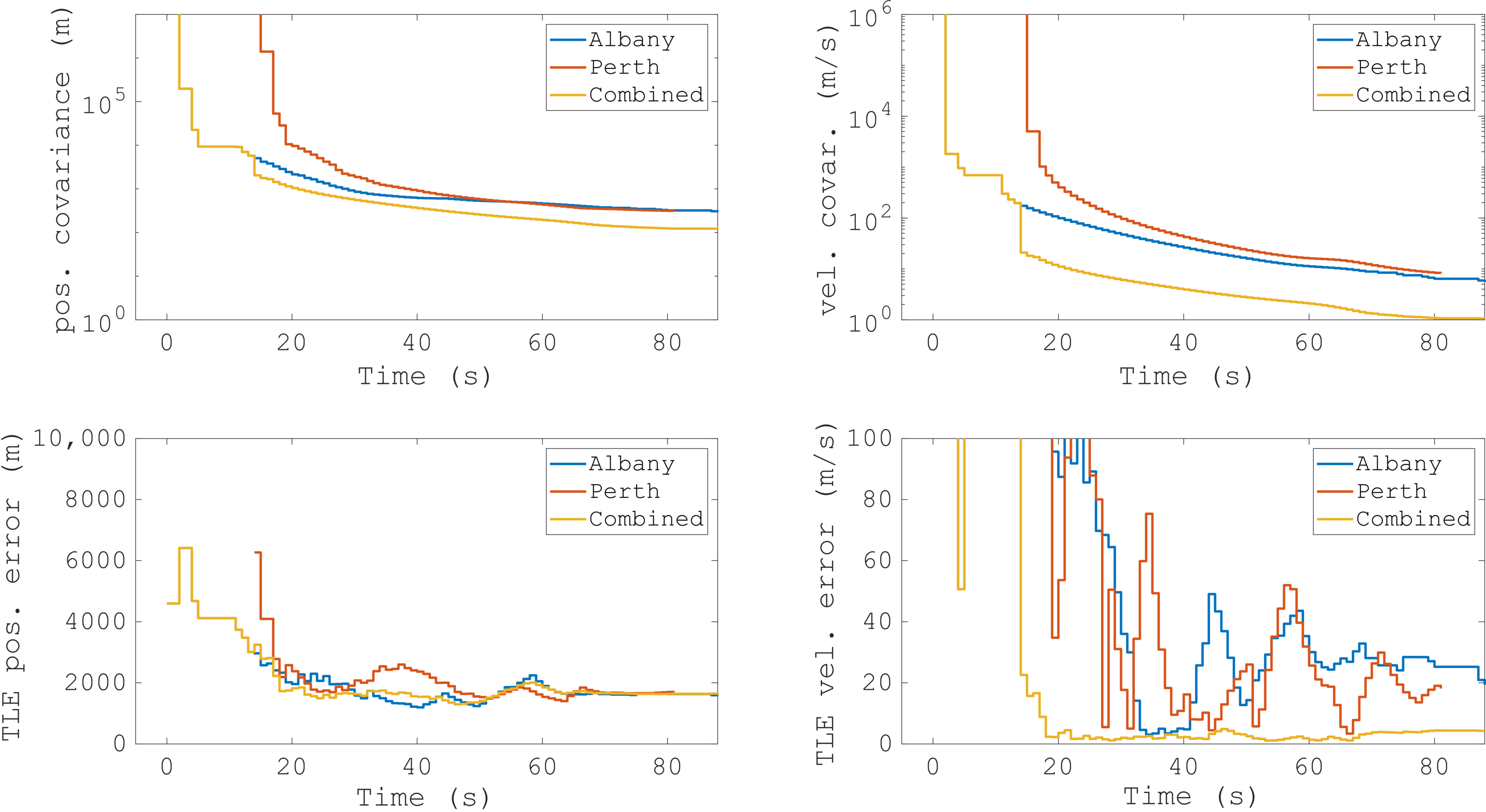}
\end{center}
\caption{The resulting orbit predictions from the multistatic measurements from {Figure} \ref{fig:OD_summary_COSMOS_1707_COMBINED_measurements}. The top panels show the covariance of the position estimate and the velocity estimate. The bottom panels show the mean errors when compared with the two-line element (TLE).}
\label{fig:OD_summary_COSMOS_1707_COMBINED_orbit_prop}
\end{figure}

Figure \ref{fig:OD_summary_COSMOS_1707_COMBINED_orbit_prop} shows the accuracy of the combined orbit, equivalent to Figure \ref{fig:OD_summary_COSMOS_1707_albany_orbit_prop} showing a single bistatic case. The combined orbit is significantly more accurate than either individual bistatic pairs, particularly the determined velocity.  A single detection from Perth (at the 17 s mark) reduces the velocity covariance by an order of magnitude, matching the expectations outlined in Section \ref{ssec:multistatic_od}.

\begin{figure}[ht!]
\begin{center}
\includegraphics[width=\columnwidth]{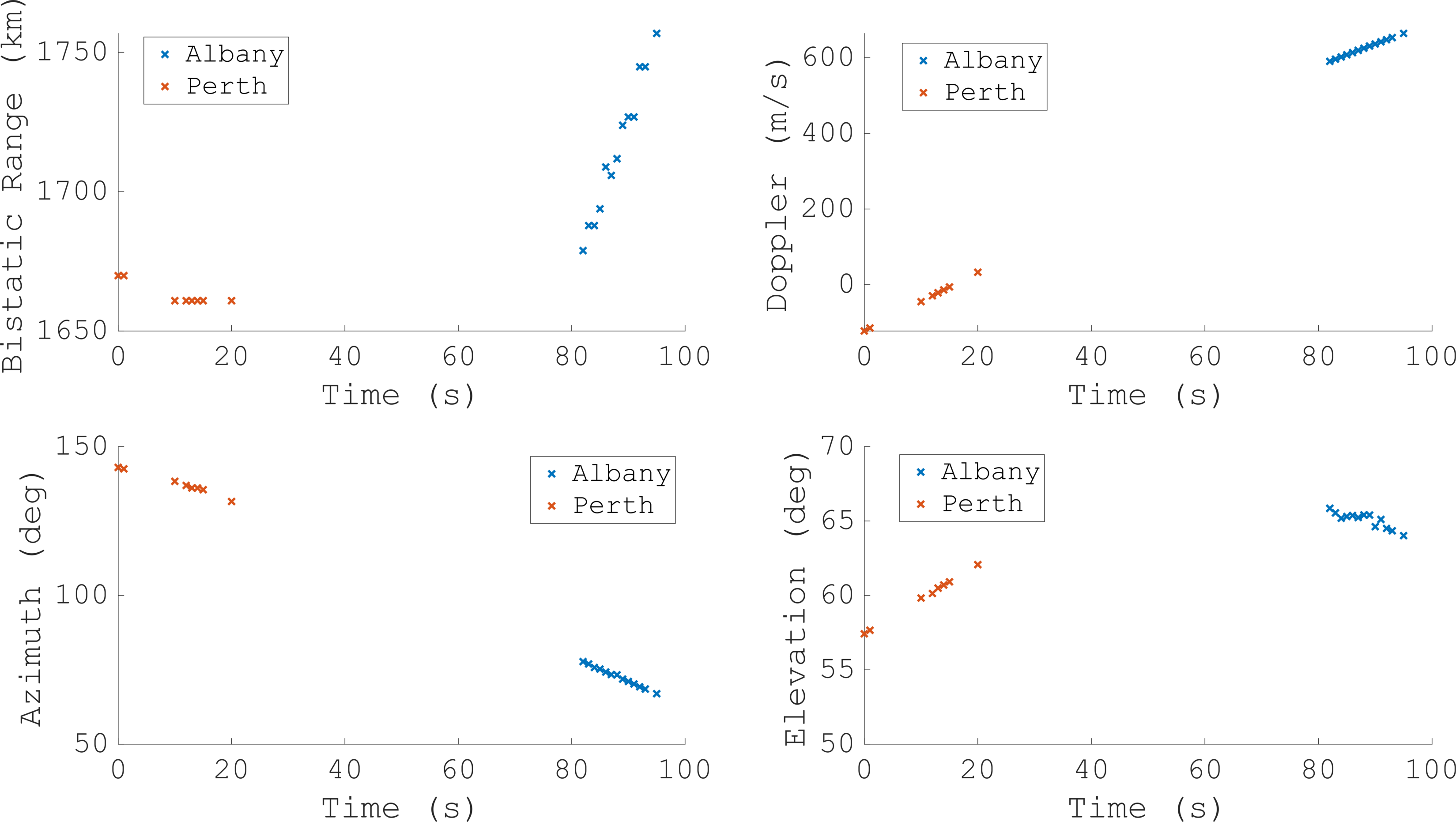}
\end{center}
\caption{The four measurement parameters from the detections of NADEZHDA 5 detected using FM transmitters in both Albany and Perth.}
\label{fig:OD_summary_NADEZHDA4_25567_COMBINED_measurements}
\end{figure}

\begin{figure}[ht!]
\begin{center}
\includegraphics[width=\columnwidth]{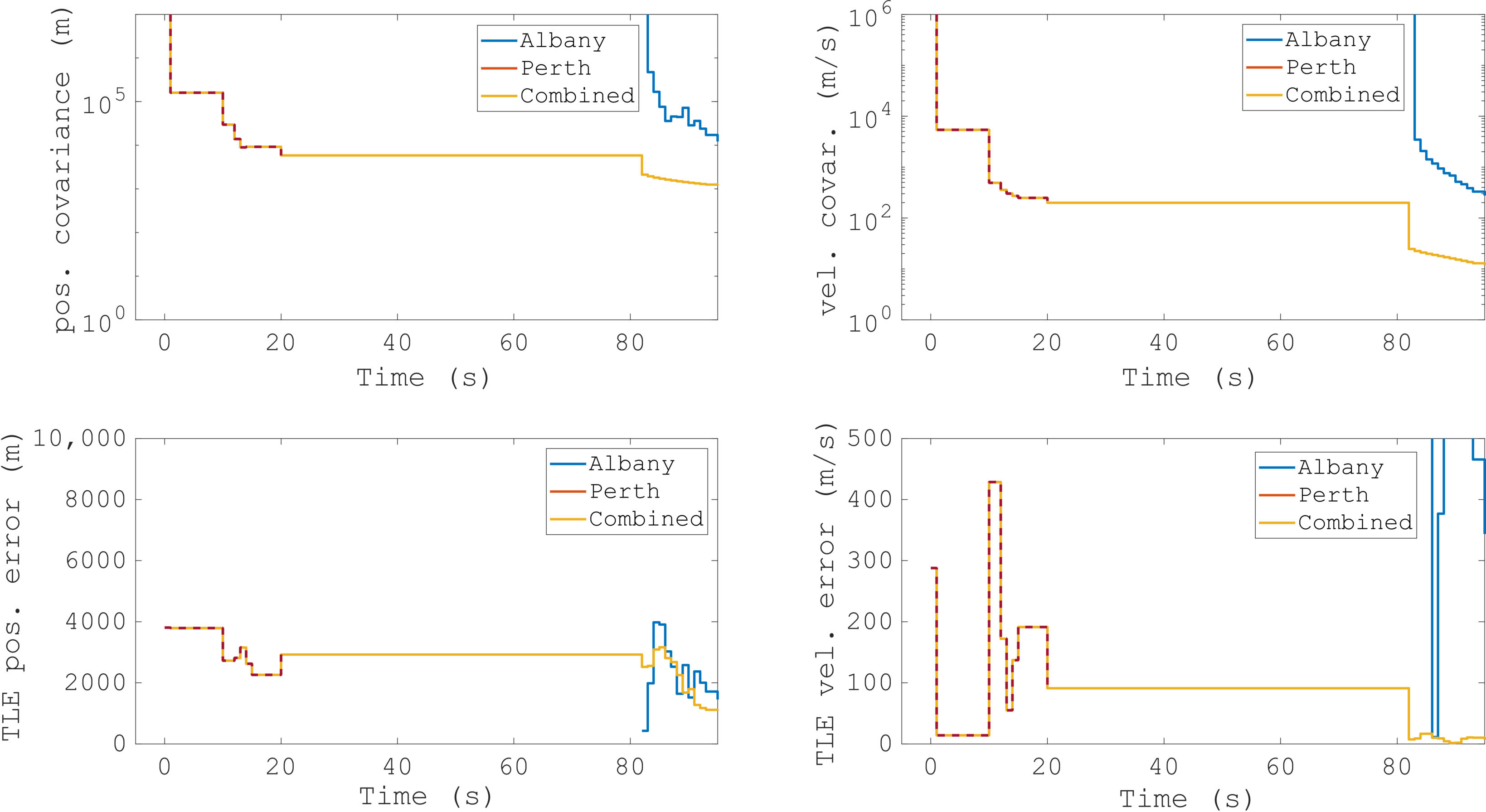}
\end{center}
\caption{The resulting orbit predictions from the multistatic measurements from {Figure} \ref{fig:OD_summary_NADEZHDA4_25567_COMBINED_measurements}. The top panels show the covariance of the position estimate and the velocity estimate. As the detections from each bistatic pair are not coincident, the combined errors will be initially identical to Perth's. The bottom panels show the mean errors when compared with the two-line element (TLE).}
\label{fig:OD_summary_NADEZHDA4_25567_COMBINED_orbit_prop}
\end{figure}

As mentioned earlier, multistatic detections do not need to be coincident to improve the overall orbit. Figures \ref{fig:OD_summary_NADEZHDA4_25567_COMBINED_measurements} and \ref{fig:OD_summary_NADEZHDA4_25567_COMBINED_orbit_prop} show the results from the detections of NADEZHDA 5 (NORAD 25567), a far smaller (albeit still classified as large) RSO at a range of 1000 km. The figures again show detections from both the Albany and Perth bistatic pairs, but instead of being coincident, the set of the detections are separated by over a minute. However, just as before, the multistatic detections greatly improve the accuracy of the orbit, confirmed both by the reduced covariance as well as compared to the TLE.

\begin{figure}[ht!]
\begin{center}
\includegraphics[width=\columnwidth]{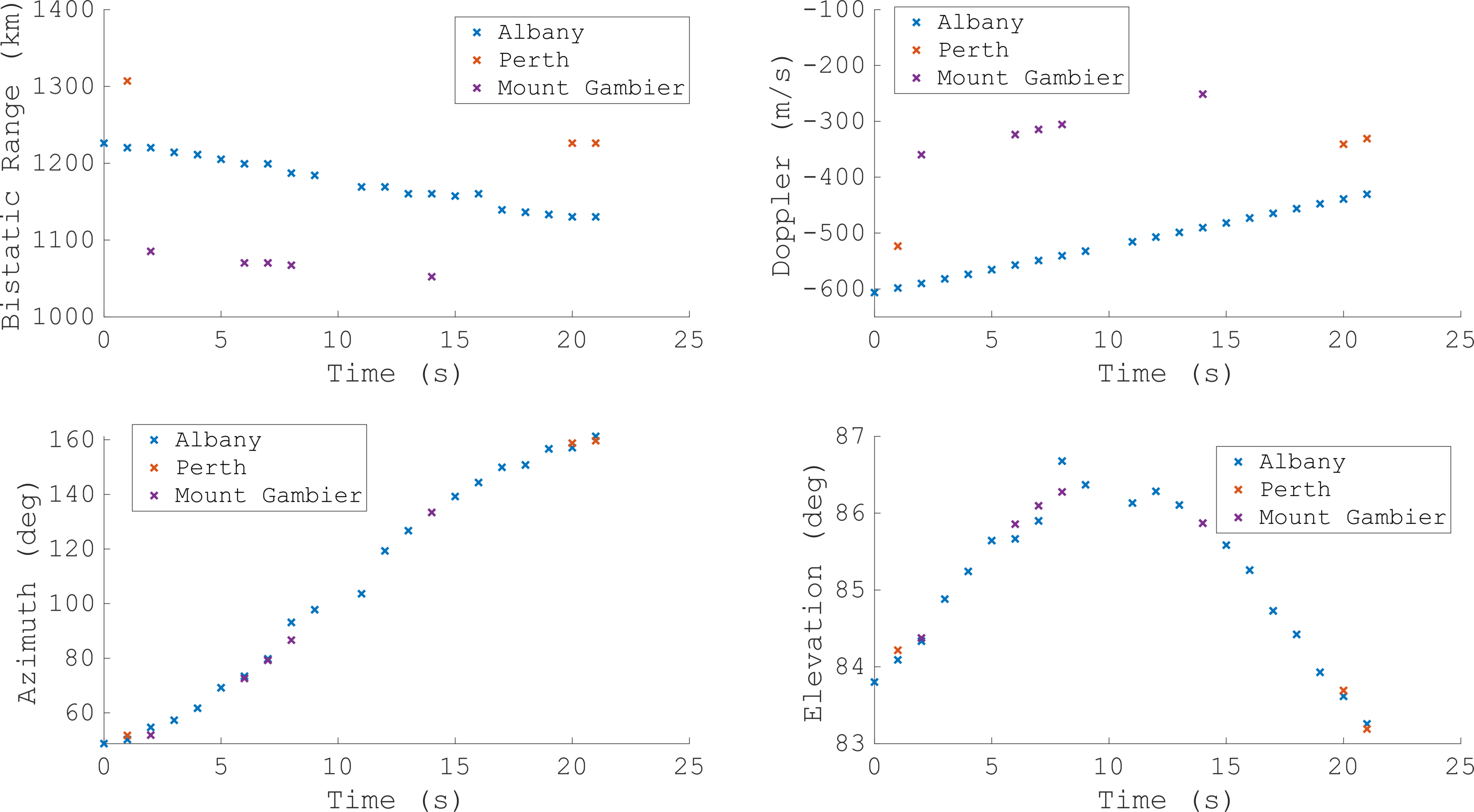}
\end{center}
\caption{The four measurement parameters from detections of OPS 5721 detected using FM transmitters in Albany, Perth and Mount Gambier.}
\label{fig:9415_OPS_5721_detections}
\end{figure}

\begin{figure}[ht!]
\begin{center}
\includegraphics[width=\columnwidth]{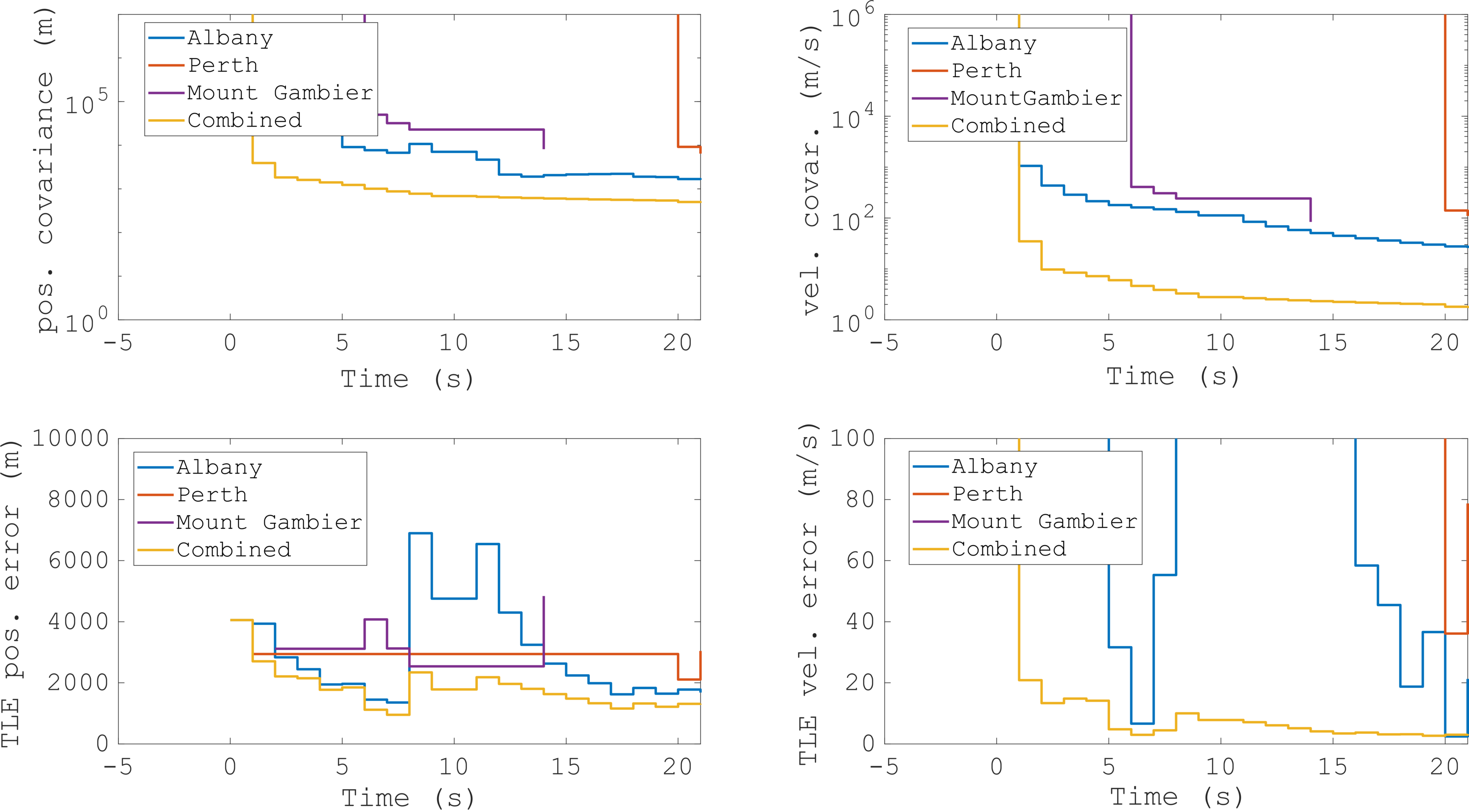}
\end{center}
\caption{{The} resulting orbit predictions from the multistatic measurements from {Figure} \ref{fig:9415_OPS_5721_detections}. The top panels show the covariance of the position estimate and the velocity estimate. The bottom panels show the mean errors when compared with the two-line element (TLE).}
\label{fig:9415_OPS_5721_OD_results}
\end{figure}

{An example of the three transmitters is simultaneously shown in Figures \ref{fig:9415_OPS_5721_detections} and \ref{fig:9415_OPS_5721_OD_results}. In this example, the satellite OPS 5721 (NORAD 9415) is detected for approximately 20 s with the Albany illuminator; however, these detections are supplemented by a small number of detections achieved utilising the Perth and Mount Gambier illuminators. Despite the short period of detections, the resulting orbit is very accurate when compared against the TLE. {Indeed, after only five seconds, the resulting orbit utilising detections across all three transmitters is very accurate.}}

There are complications when comparing and assessing determined orbits, especially when comparing them to the TLEs. Looking at the covariance of the multistatic results in \mbox{Figure \ref{fig:OD_summary_COSMOS_1707_COMBINED_orbit_prop}}, the increasing number of detections improves the estimate, especially for velocity. However, when compared to the TLE, the error does not improve; rather, it plateaus. This could be due to many factors; however, these results are within the accuracy of the TLEs themselves, as positional errors generally vary from a minimum error of approximately 1~km at the TLE's epoch up to 5~km, depending on the age of the TLE~ \cite{vallado2006revisiting,ly2020correcting}. These uncertainties could potentially mask any systematic biases or offsets, either from the system itself or from the ionosphere~\cite{hapgood2010ionospheric,holdsworth2020low}. Longer surveillance campaigns are needed to properly assess any potential systemic issues and to fully evaluate the accuracy of short-arc orbit determination.

The true RCS sizes of objects are challenging to estimate, particularly for a passive radar. Without knowing the precise details of transmitter characteristics, the amount of incident power is not known. Additionally, bistatic RCS is typically a complicated function and without accurate knowledge of the precise size and attitude of RSOs, the bistatic RCS is difficult to determine. 

The RCS values for known RSOs can be coarsely estimated with simple shapes, such as cylinders. Comparing the estimates of the detected objects provides additional data points to match against the earlier performance predictions. For example, the medium RCS satellite OV1-5 (NORAD 2122) was detected at a range of 1150 km from the MWA ({the} corresponding bistatic range was 1600 km) with an SNR of 21 dB. The maximum monostatic RCS of a cylinder of matching dimensions (1.387~m length and 0.69~m diameter) is approximately 0.7~m$^2$. For OV1-5, the RSO's length is less than half a wavelength, meaning that for smaller RSOs, the scattering will cause the RCS to decrease rapidly~\cite{knott2004radar}. Conversely, for RSOs that possess trailing antennas that have low RCS at high frequencies, these structures can produce a large RCS at MWA frequencies.  Examples such as OV1-5 agree strongly with the initial predictions that the MWA, used as a passive radar, is able to detect objects with an RCS of 0.5~m$^2$ to a range of 1000~km~\cite{2013AJ....146..103T}.

\section{Conclusion} \label{sec:conclusion}

{This paper has described the use of the MWA as a passive radar for the surveillance of space with FM radio illumination. The MWA's high time-resolution and receiving capabilities have been described, and the orbital-specific signal processing methods to form radar products have been detailed, from pulse compression through to forming detections. These orbital-specific methods are required to track an RSO’s motion throughout long CPIs to increase SNR. Following detection, the paper details the orbit determination methods, and how multistatic detections can greatly improve the orbital estimate. To demonstrate and verify these methods, this paper includes the results of a short collection campaign, utilising four transmitters across the country. With the data collected during this campaign the MWA was able to detect and accurately track every large object that passed through its main beam at a range of 1000~km or less. It additionally tracked many other objects outside these limits. These results are in agreement with earlier predictions made of the space surveillance capabilities of the MWA when used as a passive radar receiver.}

{Multiple transmitters were used to form a multistatic radar network, with multistatic detections allowing for rapid and accurate orbit determination, with additional transmitters also providing greater coverage and resilience.} By utilising these many transmitters, the MWA is able to provide persistent and widefield coverage of satellites in low Earth orbit. The MWA is able to achieve this coverage using scalable and efficient signal processing matching the radar processing to the RSO's orbit. The widefield coverage ensures that RSOs are tracked for sufficient time to accurately determine an orbit from a single pass.

{{Although it is not the SKA's intended purpose,} using the techniques described here, the SKA will be a highly capable space surveillance sensor when used as a radar~\cite{6652046}. {The SKA will be significantly more sensitive than the MWA, and even a modest increase in sensitivity would enable the detection of signals from a higher altitude (for example, from 1000~km to 2000~km). As shown in Figure \ref{fig:power_on_orbit}, higher altitudes would intrinsically allow the utilisation of transmitters at even greater distances, including along Australia's eastern seaboard.} Conversely, since transmissions reflected from RSOs are a source of interference for astrophysical observations, knowledge of which RSOs are above the horizon will be important information for the removal of RSO effects from SKA observations. {Incorporating a system to remove interference reflected from RSOs would be a natural extension for any radio telescope space surveillance radar.}}

\section{Acknowledgement}
This scientific work makes use of the Murchison Radioastronomy Observatory, operated by CSIRO. We acknowledge the Wajarri Yamatji people as the traditional owners of the Observatory site. Support for the operation of the MWA is provided by the Australian Government, under a contract to Curtin University administered by Astronomy Australia Limited. We acknowledge the Pawsey Supercomputing Centre which is supported by the Western Australian and Australian Governments.

The authors would like to thank David Holdsworth and Nick Spencer for their invaluable feedback and reviews of the manuscript. 

This work is supported by the Defence Science and Technology Group.
\bibliographystyle{IEEEtran}
\bibliography{mainbib}

\end{document}